\documentclass[journal]{IEEEtran}
\usepackage{amsmath,amsfonts}
\usepackage{algorithmic}
\usepackage{algorithm}
\usepackage{array}
\usepackage[caption=false,font=normalsize,labelfont=sf,textfont=sf]{subfig}
\usepackage{textcomp}
\usepackage{stfloats}
\usepackage{url}
\usepackage{verbatim}
\usepackage{graphicx}
\usepackage{cite}

\usepackage{multirow}
\usepackage[table]{xcolor}
\usepackage{amsmath}
\usepackage{amsthm}
\usepackage{booktabs}
\usepackage{hyperref}

\begin{document}

\title{Quantization Aware Attack: Enhancing Transferable Adversarial Attacks by Model Quantization}

\author{Yulong Yang,
        Chenhao Lin$^*$,~\IEEEmembership{Member,~IEEE,} 
        Qian Li,~\IEEEmembership{Member,~IEEE,} 
        Zhengyu Zhao,~\IEEEmembership{Member,~IEEE,}
        Haoran Fan,
        Dawei Zhou, 
        Nannan Wang,~\IEEEmembership{Member,~IEEE,} 
        Tongliang Liu,~\IEEEmembership{Senior Member,~IEEE,}
        Chao Shen,~\IEEEmembership{Senior Member,~IEEE,}
\thanks{Yulong Yang, Chenhao Lin, Qian Li, Zhengyu Zhao, Haoran Fan, and Chao Shen are with the School of Cyber Science and Engineering, Xi'an Jiaotong University, Xi'an, China.}
\thanks{Dawei Zhou, and Nannan Wang are with Xidian University, Xi'an, China.}
\thanks{Tongliang Liu is with The University of Sydney, Sydney, Australia.}
\thanks{Corresponding author: Chenhao Lin (linchenhao@xjtu.edu.cn)}
}

\markboth{Journal of \LaTeX\ Class Files,~Vol.~14, No.~8, August~2021}%
{Shell \MakeLowercase{\textit{et al.}}: A Sample Article Using IEEEtran.cls for IEEE Journals}


\maketitle

\begin{abstract}
Quantized neural networks (QNNs) have received increasing attention in resource-constrained scenarios due to their exceptional generalizability. However, their robustness against realistic black-box adversarial attacks has not been extensively studied.
In this scenario, adversarial transferability is pursued across QNNs with different quantization bitwidths, which particularly involve unknown architectures and defense methods. Previous studies claim that transferability is difficult to achieve across QNNs with different bitwidths on the condition that they share the same architecture.
However, we discover that under different architectures, transferability can be largely improved by using a QNN quantized with an extremely low bitwidth as the substitute model. We further improve the attack transferability by proposing \textit{quantization aware attack} (QAA), which fine-tunes a QNN substitute model with a multiple-bitwidth training objective.
In particular, we demonstrate that QAA addresses the two issues that are commonly known to hinder transferability: 1) quantization shifts and 2) gradient misalignments.
Extensive experimental results validate the high transferability of the QAA to diverse target models.
For instance, when adopting the ResNet-34 substitute model on ImageNet, QAA outperforms the current best attack in attacking standardly trained DNNs, adversarially trained DNNs, and QNNs with varied bitwidths by 4.3\% $\sim$ 20.9\%, 8.7\% $\sim$ 15.5\%, and 2.6\% $\sim$ 31.1\% (absolute), respectively.
In addition, QAA is efficient since it only takes one epoch for fine-tuning.
In the end, we empirically explain the effectiveness of QAA from the view of the loss landscape. 
Our code is available at ~\url{https://github.com/yyl-github-1896/QAA/}.
\end{abstract}


\section{Introduction}

Recent progress in edge computing~\cite{liu2022bringing} has triggered a large demand for deep neural networks (DNNs) that can be deployed on edge devices. As a popular DNN compression and acceleration technique, quantization is widely adopted to reduce computational overhead by replacing 32-bit full-precision DNN parameters (a.k.a. weights) with lower numbers of bitwidths (e.g., 8, 5, 4, 3, 2, and 1-bit settings) while still maintaining high accuracy.

However, many works have shown that DNNs are vulnerable to adversarial examples ~\cite{szegedy2013intriguing,guo2022specpatch,pang2020tale,papernot2016limitations,calzavara2019adversarial,ma2023transferable,zhao2020towards}, which are crafted by adding human-imperceptible perturbations and can mislead DNNs. Although the robustness of full-precision DNNs has been extensively studied, few works have focused on the robustness of QNNs, 
especially from the perspective of realistic black-box settings.
A natural question is as follows: \textit{Can we perform realistic black-box attacks against DNNs with unknown architecture and quantization bitwidths?}

\begin{figure}
    \centering
    \includegraphics[scale=0.27]{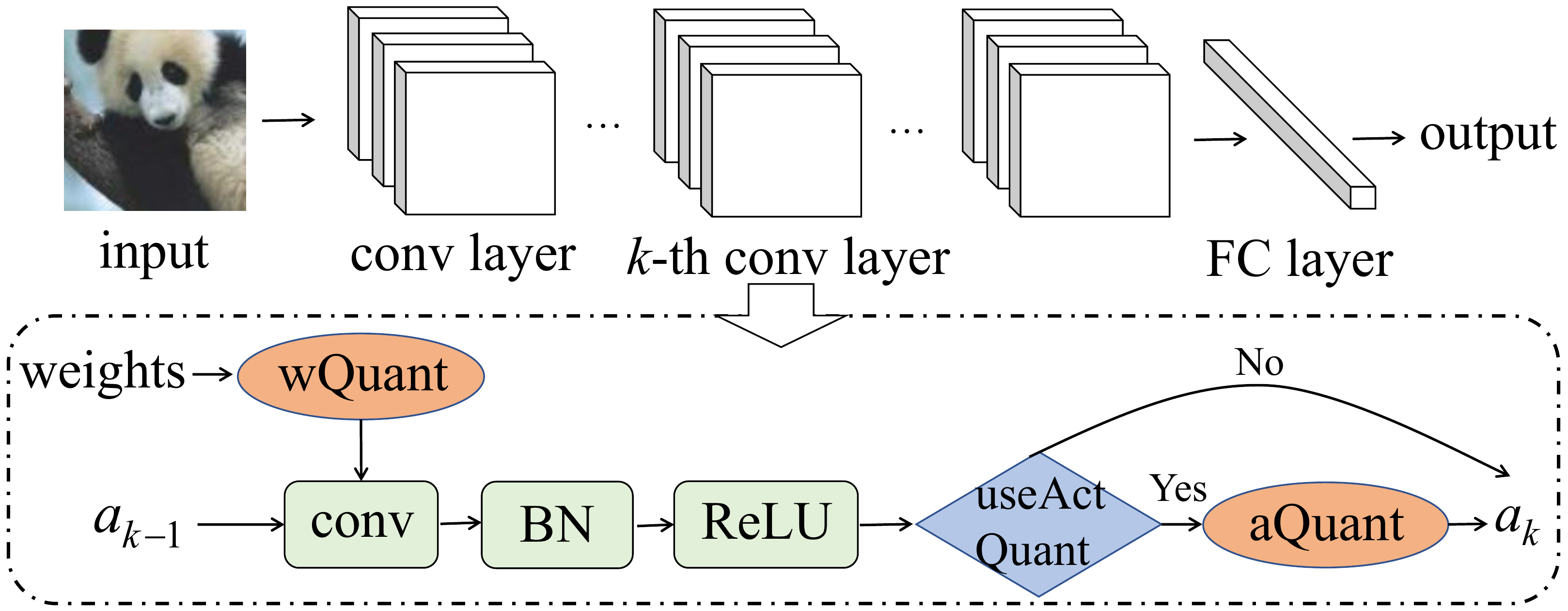}
    \captionsetup{justification=centering}
    \caption{A brief overview of the QAA substitute model.}
    \label{fig:qaa}
\end{figure}
One intriguing property of adversarial examples is their transferability across different DNNs~\cite{szegedy2013intriguing,dong2023restricted,zhu2023multi,zhong2020towards,wang2023towards}, enabling an adversary to generate adversarial examples on a local substitute model and use them to attack the target model~\cite{papernot2017practical}. The adversarial transferability property motivates us to achieve black-box attacks against unknown QNNs through transfer-based attacks. The main challenge is that many previous studies~\cite{bernhard2019impact,sen2020empir,fu2021double} have claimed that model quantization makes QNNs more robust. Specifically, the widely-held claim~\cite{bernhard2019impact} states that adversarial examples transfer poorly across QNNs with different quantization bitwidths because of the following two issues. 1) \textit{Quantization shift}: Quantization can map two different feature values (i.e. clean feature values and adversarial feature values) into the same bucket, ruining the adversarial effect. 2) \textit{Gradient misalignment}: Quantization approximately computes gradients with the straight-through estimator (STE)~\cite{bengio2013estimating}, making the gradients of QNNs greatly diverge from that of full-precision DNNs.

However, we discover that the above claim of ``poor transferability'' is only valid when the substitute model and the target model have the same architecture. When the architectures of the substitute model and the target model differ, QNN substitute models with ``extremely low bitwidths'' (less than 5 bits) may become the ``panacea'' for attacking an unknown target model (see Sec.~\ref{subsec:preliminaries} for more details). Please note that the ``different architecture'' setting is closer to the realistic black-box scenario, and our new finding indicates that an adversary can leverage model quantization to generate highly transferable adversarial examples. However, technically, the attack capability of the QNN substitute model is still unsatisfactory because of the quantization shift and the gradient misalignment issues.
The following question now arises: \textit{Can we mitigate the above two issues to improve the transferability of adversarial examples across QNNs with different bitwidths?}

To enhance the attack transferability, a naive approach is to directly ensembles substitute models with different quantization bitwidth (a.k.a. ensemble-based attack). However, an ensemble-based attack multiplies the computational overhead and leads to a suboptimal attack success rate (for additional details, see Sec.~\ref{subsec:ablation}). To this end, we propose a quantization aware attack (QAA), which fine-tunes substitute QNNs with objective functions possessing multiple quantization bitwidths, making the substitute model ``aware of'' the target of attacking QNNs with unknown bitwidths. Fig.~\ref{fig:qaa} presents a brief overview of the QAA substitute model, which alternatively utilizes quantized activation and full-precision values during the forward propagation process to promote adversarial transferability. The advantage of the QAA is its ability to introduce gradients of QNNs with different bitwidths during the substitute model training stage and utilize various bitwidths activation values at the inference stage, thus overcoming the quantization shift and gradient misalignment challenges and improving the transferability of the generated adversarial examples, as shown in Sec.~\ref{sec:method}.

We demonstrate the effectiveness of the proposed QAA on both the CIFAR-10 and ImageNet datasets. The experimental results show that compared with the state-of-the-art (SOTA) attacks which were designed only for full-precision target models, the QAA can significantly enhance the transferability of adversarial examples to standardly trained DNNs, adversarially trained DNNs, and QNNs with varied bitwidths by 4.6\% $\sim$ 20.9\%, 8.8\%  $\sim$ 13.4\% and 2.6\%  $\sim$ 11.8\% (absolute), respectively. Finally, we attempt to understand the underlying causes of the enhancements made by the QAA from the loss landscape perspective~\cite{keskar2016large}, in which the explained numerical results are consistent with the attack performance. 
In summary, our main contributions are as follows.
\begin{itemize}
    \item For the first time, we examine the widely-held claim that adversarial attacks transfer poorly across QNNs with different bitwidths in the black-box setting, and enhance the attack transferability by mitigating the quantization shift and gradient misalignment issues.
     \item We propose a \textit{quantization aware attack (QAA)}, which trains substitute models with multiple bitwidths objectives and endows the models with multiple bitwidths awareness capability to solve the above problems.
    \item Comprehensive evaluations demonstrate the improvement yielded by the proposed QAA over several other SOTA attacks. Numerical evidence derived from the loss landscape view further verifies the effectiveness of QAA.
\end{itemize}

\section{Related Work}

\subsection{Transfer-based Black-box Attacks}
\textcolor{black}{Papernot et al.~\cite{papernot2017practical} first proposed the concept of transfer-based black-box attacks, in which the adversary trains a local substitute model to obtain transferable adversarial examples. Various methods have been developed to enhance the transferability of adversarial examples.
These attack methods can be divided into five categories~\cite{zhao2023revisiting}: momentum-based attacks~\cite{dong2018boosting,lin2020nesterov,wang2021enhancing,xiong2022stochastic,weng2023logit}, input-transformation-based attacks~\cite{xie2019improving,wang2021admix,long2022frequency},  feature-level attacks~\cite{huang2019enhancing,wang2021feature,zhao2021success,zhang2022improving}, GAN-based attacks~\cite{salzmann2021learning,wang2023towards}, and model-based attacks~\cite{wu2020skip,gubri2022lgv,zhu2022toward}.}
Momentum-based attacks add regularization terms to the original attack objectives when attacking a substitute model; thus, the optimization gradient can be stabilized and the attack transferability can be enhanced. Input-transformation-based attacks mimic model ensembles by applying image-transformation operations when attacking the substitute model to achieve enhanced transferability. Feature-level attacks operate under the hypothesis that the features of the shallower DNN layers represent basic and invariant image information and, thus, are more significant for generating transferable adversarial examples. ~\textcolor{black}{GAN-based attacks leverage generative models to obtain transferable adversarial examples, and they are mighty for targeted attacks~\cite{wang2023towards}. Model-based attacks devise a substitute model to smooth the loss landscape and thus enhance attack transferability. }
This paper describes a model-based attack, in which the transferability of adversarial examples across QNNs with different architectures and bitwidths is investigated by quantizing and fine-tuning the substitute model.

\subsection{Attacks and Defenses for QNNs}
Adversarial attacks can be divided into white-box attacks and black-box attacks. Existing works have only designed white-box attacks against QNNs~\cite{gupta2020improved}. 
However, how to improve black-box attacks against both full-precision DNNs and QNNs with different architectures and quantization bitwidths is still an open problem. The adversarial robustness of QNNs has received increased research attention from the defender's perspective. For example, existing works have designed adversarial training~\cite{fu2021double}, regularization-based defense~\cite{song2020improving}, and ensemble-based defense~\cite{sen2020empir} strategies for QNNs. In particular, previous works have shown adversarial examples transfer poorly across DNNs with the same architecture but different quantization bitwidths because of the quantization shift and gradient misalignment phenomena~\cite{bernhard2019impact}. Intuitively, this characteristic of QNNs makes it difficult for the adversary to successfully attack QNNs with different bitwidths simultaneously, enabling defenders to enhance the robustness of DNNs by ensembling QNNs with different bitwidths. For instance, EMPIR~\cite{sen2020empir} directly ensembles full-precision DNNs and low-bitwidth QNNs with the different bitwidths to enhance the robustness, and Double-Win Quant~\cite{fu2021double} alternatively applies the adaptive quantization technique to the adversarial training scheme to enable QNNs to exhibit different bitwidths during inference. The research gap lies in the fact that previous works only studied the transferability of adversarial examples under the ``same architecture, different quantization bitwidth'' setting, while the conclusion can be different under the ``different architecture, different quantization bitwidth'' setting, which is closer to the realistic black-box scenario.
In summary, there are two dimensions along which our work differs from the others. The first is that it exhibits improved adversarial transferability across target models with unknown architectures and bitwidths. The second is that it quantizes and fine-tunes the substitute model to achieve the above goal.

\subsection{DNN Quantization}
\textcolor{black}{DNN quantization compresses and accelerates DNNs by representing network weights/activations/gradients with lower bitwidth. Model quantization has been widely applied in efficient DNN inference frameworks in industry~\cite{jacob2018quantization}. Two main categories of quantization techniques are quantization aware training (QAT)~\cite{hubara2016binarized,rastegari2016xnor,zhou2016dorefa,li2019additive} and post training quantization (PTQ)~\cite{nagel2020up,li2021brecq,hubara2021accurate,wei2022qdrop}.} QAT trains DNNs from scratch or fine-tunes full-precision DNNs given a training dataset $\mathcal{D}_{train} = \{x_i, y_i\}_{i=1}^n$ to convert these DNNs into low-bitwidth DNNs. In contrast, PTQ does not require end-to-end training and instead relies on a calibration dataset $\mathcal{D}_{cali} = \{x_i, y_i\}_{i=1}^m$ ($m<<n$) to implement a rounding scheme. PTQ is more computationally efficient than QAT, but it has greater quantization loss. Recent PTQ-related works~\cite{nagel2020up,li2021brecq,hubara2021accurate,wei2022qdrop} have modeled weight/activation quantization as perturbations and used Taylor expansion to analyze loss value changes before reconstructing the output of each layer. To implement QAT or PTQ algorithms on GPUs, one can use fake quantization to mimic QNNs, in which the weights are still stored at full-precision but are rounded into lower bitwidth during inference. QAT process with fake quantization can be formalized as follows,
\begin{equation}
 \min_{w, \beta} \limits {\quad \sum_{i=1}^n {\mathcal{L}(f(x_i,w,\beta),y_i)}},
\end{equation}
in which $\mathcal{L(\cdot)}$ denotes the loss function (e.g., cross-entropy loss), $w$ is the DNN weight, and $\beta$ is the quantization hyper-parameter used to convert $w$ from full-precision floating points into lower bitwidth numbers. PTQ with fake quantization can be formalized as follows.
\begin{equation}
 \min_{\beta} \limits {\quad \sum_{j=1}^m {\mathcal{L}(f(x_j,w,\beta),y_j)}}.
\end{equation}
The major difference between QAT and PTQ is that the PTQ only optimizes quantization hyper-parameter $\beta$ while keeping the model weights $w$ unchanged.
\textcolor{black}{Apart from the above model quantization techniques with fixed bitwidths, a line of work has studied adaptive bitwidth quantization~\cite{jin2020adabits,guerra2020switchable,fu2021double}, which can switch the quantization bitwidth after deployment. The application of such adaptive quantization techniques in realistic settings further urges us to develop effective transfer-based attacks against QNNs with different bitwidth.}

\section{Methodology}
\label{sec:method}
In this section, we first show our motivating observation, that is, extremely low bitwidth quantized substitute models benefit the generation of transferable adversarial examples. Next, we formulate the black-box transfer attack problem and present our QAA fine-tuning objective function to achieve the attack goal. Finally, we provide a numerical analysis of how the proposed QAA method mitigates the quantization shift and gradient misalignment issues and further enhances the attack transferability of quantized substitute models.

\begin{table*}[]
\centering
\caption{Transfer Attack Success Rates (\%) Across QNNs With Different Bitwidths (32, 8, 5, 4, 3, 2-bit) And Architectures (ResNet-18 and ResNet-34) On ImageNet.
}
\begin{tabular}{c|c|ccccccc|ccccccc}
\toprule
                        \multirow{2}{*}{Substitute} & \multirow{2}{*}{Bitwidth} & \multicolumn{7}{c}{ResNet-18}                   & \multicolumn{7}{c}{ResNet-34}                    \\
                        &                           & 32   & 8   & 5      & 4      & 3      & 2   & Avg.    & 32   & 8   & 5      & 4      & 3       & 2 & Avg.       \\
\midrule
\multirow{2}{*}{ResNet-18} & 32                        & 100.0 & 99.0 & 90.3  & 88.4  & 88.3  & 85.4  & 91.9   & 83.3  & 82.1  & 76.8  & 78.0  & 76.8   & 78.2  & 79.2 \\
                        & 2                         & 91.8  & 92.1 & 99.2  & 99.7  & 99.8  & 100.0  & 97.1 & 80.8  & 81.0  & 79.5  & 82.8  & 84.8   & 89.2  & 83.0 \\
\multirow{2}{*}{ResNet-34} & 32                        & 89.7  & 88.3  & 86.0  & 85.6  & 83.8  & 80.6  & 85.7  & 100.0 & 99.2 & 78.8  & 79.6  & 79.6   & 76.7  & 85.7 \\
                        & 2                         & 91.9  & 92.4  & 92.8  & 94.3  & 94.1  & 95.3  & 93.5  & 87.3 & 89.0  & 98.1  & 99.0  & 100.0  & 100.0 & 95.6 \\
\bottomrule
\end{tabular}
\label{tab:preliminaries}
\end{table*}

\subsection{Preliminaries: Transfer-based Attacks against QNNs}
\label{subsec:preliminaries}
In the realistic black-box scenario, the architecture and quantization bitwidth of the target model are unknown to the adversary. Thus, it is important to study the transferability of adversarial examples across DNNs with different architectures and quantization bitwidths. However, previous works claimed that adversarial examples transfer poorly across DNNs with different quantization bitwidths~\cite{bernhard2019impact,sen2020empir,fu2021double}. To verify this claim, we conduct illustrative transfer-based attack experiments on the ImageNet dataset with two model architectures (ResNet-18, ResNet-34) and six quantization bitwidths (32, 8, 5, 4, 3, and 2-bit settings). The model quantization technique is APoT~\cite{li2019additive}, and the transfer-based attack algorithm is MIM~\cite{dong2018boosting} with a $l_{\infty}$-norm perturbation budget $\epsilon=16/255$. The experimental results are shown in Tab.~\ref{tab:preliminaries}. From these results, we have the following observations.

\textbf{Observation 1: Model quantization harms the same-architecture transferability of adversarial examples.} Given that the substitute model has the same architecture as that of the target model, the use of different quantization bitwidth lowers the transfer attack success rate. For instance, when using ResNet-18 with 32-bit as the substitute model, the attack success rate decreases as the quantization bitwidth of the target model decreases; when using ResNet-18 with 2-bit as the substitute model, the attack success rates increase as the quantization bitwith of the target model decreases. This observation is exactly the same as the previous ``poor transferability'' conclusion~\cite{bernhard2019impact,sen2020empir,fu2021double}.

\textbf{Observation 2: Model quantization benefits the cross-architecture transferability of adversarial examples.} The conclusion can be different when the substitute model and the target model have different architectures. For instance, when taking ResNet-34 as the substitute model to attack the ResNet-18 target model, the 2-bit substitute model outperforms the 32-bit model under every target bitwidth. A similar phenomenon occurs when ResNet-18 is used as the substitute model to attack ResNet-34. Although the 2-bit substitute model does not achieve higher attack success rates on the 32-bit target model, it outperforms the 32-bit substitute model at other target bitwidths and has higher average attack success rates.

\textbf{Remark on the above observations.} The same-architecture observation (Observation 1) indicates that the defender may ensemble models with different quantization bitwidths to hinder the transferability of adversarial examples and enhance the adversarial robustness of their models; this strategy has been explored in previous works~\cite{sen2020empir,fu2021double}. The cross-architecture observation (Observation 2) indicates that model quantization may enhance the transfer attack ability of the adversary under black-box scenarios, which was ignored by previous works. Please note that the setting of Observation 2 is closer to the realistic black-box setting where both the target architecture and the quantization bitwidth are unknown to the adversary. Given that the adversary has no portable way to detect the architecture or the bitwidth of the target model, the best choice for the adversary is to leverage 2-bit quantized substitute models to achieve the best average attack performance. In the next subsection, we discuss how to achieve the best black-box attack capability from the perspective of the adversary and propose our QAA method.

\subsection{Problem Formulation}
Quantization compresses and accelerates DNNs by converting the 32-bit weight and activation values ($w$ and $a$) into lower bitwidths ($\hat{w}$ and $\hat{a}$):
\begin{equation}
\label{eqn:quantization}
    \begin{gathered}
        \hat{w} = \text{Int}(\frac{w-b_w}{s_w}),\  
        \hat{a} = \text{Int}(\frac{a-b_a}{s_a}),
    \end{gathered}
\end{equation}
where $b_w, s_w, b_a, s_a$ are the bias and scaling terms for weight quantization and activation quantization, respectively; Int denotes the integer operation.
The adversary's goal is to maximize the probability of fooling target model $f$ with unknown architecture $m$ and bitwidths $q$, that is, 
\begin{equation}
\label{attack_objective}
    \begin{gathered}
         \max_{\delta} \sum_{q} \sum_{m} \mathcal{P}(f_{q, m} (x+\delta)\neq y | q, m) p(q) p (m),
    \end{gathered}
\end{equation}
where $x$ and $y$ are clean example and its true label, respectively, $\delta$ is the adversarial perturbation, $\mathcal{P}$ denotes the probability, and $f_{q, m}$ denotes target model $f$ with certain bitwidths $q$ and architecture $m$. 

To generate adversarial examples, the adversary needs to train a local substitute model $\overline{f}$ with certain architecture $\overline{m}$ and bitwidth $\overline{q}$ on the available training dataset and generate adversarial examples, that is,
\begin{equation}
\label{attack_substitute}
    \begin{gathered}
        \max_{\delta} \mathcal{L}(\overline{f}_{\overline{q},\overline{m}}(x + \delta), y),
    \end{gathered}
\end{equation}
where $\mathcal{L}$ is the surrogate loss function.

From Eqn.~\ref{attack_objective} and Eqn.~\ref{attack_substitute}, we can see that the difference between the architectures and bitwidths of the substitute and target model is the prominent factor that hinders the transferability of the generated adversarial examples. A previous work~\cite{bernhard2019impact} has proposed the following two hypotheses to explain the poor transferability of adversarial examples across target DNNs and QNNs with different bitwidths.

\textbf{Quantization shift.} Quantization can map two different types of feature values (i.e., clean features and adversarial features) into the same bucket, damaging the adversarial effect.

\textbf{Gradient misalignment.} QNNs approximately compute gradients with the straight-through estimator (STE)~\cite{bengio2013estimating}, making the gradients of QNNs diverge largely from those of full-precision DNNs.

\subsection{Objective Function}
This paper proposes a quantization aware attack (QAA) to address the above two challenges by making the substitute model training procedure ``aware of'' quantization objective with different bitwidths, that is,
\begin{equation}
\label{obj_function}
    \min_{w, \beta} \sum_{q} \sum_{i} \mathcal{L}(\overline{f}_q(x_{i}, w, \beta), y_{i}),
\end{equation}
where $\overline{f}_q$ denotes the substitute model with bitwidths $q$, weights $w$, and quantization parameter set $\beta$ (including the $b_w, s_w, b_a, s_a$ mentioned above). This objective function means that a single substitute model is quantized to multiple bitwidths when calculating losses and gradients, making the substitute model ``aware of'' different quantization bitwidths. At the attack iteration stage, the substitute model performs inference with diverse bitwidths to prevent adversarial examples from overfitting to attacking the model with a single bitwidth, thus enhancing the transferability of the approach.

The QAA addresses the poor transferability issue of adversarial examples across QNNs in two aspects. From the quantization shift perspective, the QAA performs multiple bitwidths inferences in the attack generation stage, forcing the algorithm to find adversarial examples that are effective on the target QNNs with multiple bitwidths. From the gradient misalignment perspective, the proposed QAA training objective Eqn.~\ref{obj_function} makes the gradients of the substitute model align with target QNNs. We provide quantitative analysis in Sec.~\ref{subsec:quantitiative_analysis_on_qaa} to support the above claims.

\subsection{Attack Implementation}
\begin{figure}
    \centering
    \includegraphics[scale=0.5]{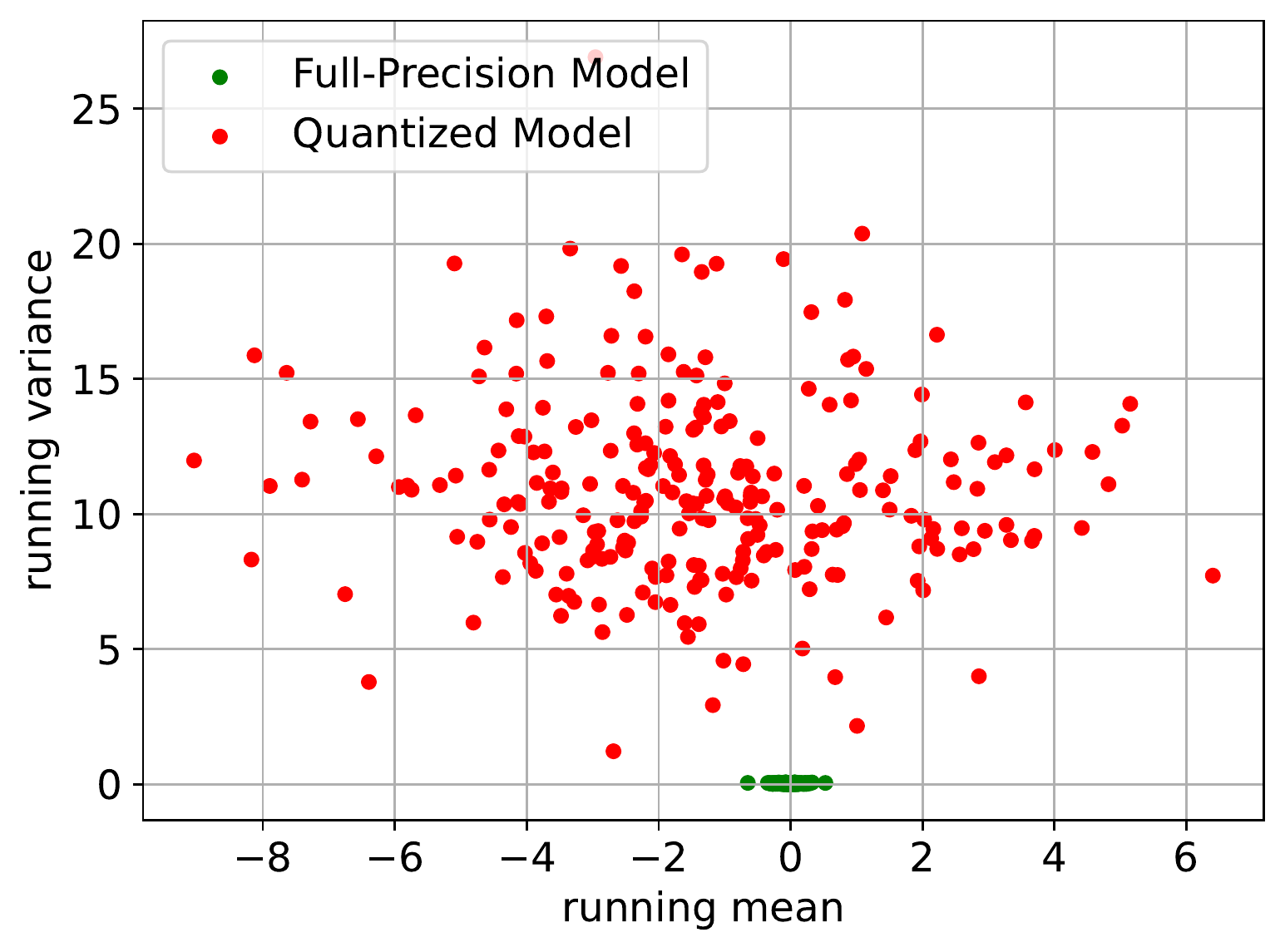}
    \caption{The BN challenge of training substitute model with multiple bitwidths.}
    \label{fig:bn_visulization}
\end{figure}
\begin{algorithm}[t]
\caption{Quantization Aware Attack based on QAT}\label{alg:qaa}
\begin{algorithmic}
\STATE \textbf{Require:} a training dataset $\mathcal{D}_{train}$, a pre-trained QNN $\overline{f}$
\STATE \textbf{a. Substitute model training:}
\STATE \hspace{0.05cm} 1: \textbf{Initialize:} $useActQuant = True$
\STATE \hspace{0.05cm} 2: \textbf{for} $(x, y) \in \mathcal{D}_{train}$ \textbf{do}
\STATE \hspace{0.05cm} 3: \hspace{0.5cm} $useActQuant$ = \textbf{not} $useActQuant$
\STATE \hspace{0.05cm} 4: \hspace{0.5cm} \textbf{if} $useActQuant$ \textbf{do}
\STATE \hspace{0.05cm} 5: \hspace{0.5cm} \hspace{0.5cm} $output = \overline{f}(x, \hat{w}, \hat{a})$
\STATE \hspace{0.05cm} 6: \hspace{0.5cm} \hspace{0.5cm} $loss = \mathcal{L}(output, y)$
\STATE \hspace{0.05cm} 7: \hspace{0.5cm} \hspace{0.5cm} update $w, b_{w}, s_{w}, b_a, s_a$
\STATE \hspace{0.05cm} 8: \hspace{0.5cm}  \textbf{else do}
\STATE \hspace{0.05cm} 9: \hspace{0.5cm}  \hspace{0.5cm} $output = \overline{f}(x, \hat{w}, a)$
\STATE 10: \hspace{0.5cm} \hspace{0.5cm} $loss = \mathcal{L}(output, y)$
\STATE 11: \hspace{0.5cm}  \hspace{0.5cm} update $w, b_{w}, s_{w}$
\STATE 12: \hspace{0.5cm} \textbf{end if}
\STATE 13: \textbf{end for}

\STATE \textbf{b. Adversarial example generation:}
\STATE 14: \textbf{Initialize:} $useActQuant = True$, $x_{adv}=x$
\STATE 15: \textbf{Repeat} for N iterations:
\STATE 16: \hspace{0.5cm} $useActQuant$ = \textbf{not} $useActQuant$
\STATE 17: \hspace{0.5cm} \textbf{if} $useActQuant$ \textbf{do}
\STATE 18: \hspace{0.5cm}  \hspace{0.5cm} $output = \overline{f}(x_{adv}, \hat{w}, \hat{a})$
\STATE 19: \hspace{0.5cm} \textbf{else do}
\STATE 20: \hspace{0.5cm} \hspace{0.5cm} $output = \overline{f}(x_{adv}, \hat{w}, a)$
\STATE 21: \hspace{0.5cm} \textbf{end if}
\STATE 22:  \hspace{0.5cm}  $loss = \mathcal{L}(output, y)$
\STATE 23: \hspace{0.5cm} update $x_{adv}$
\end{algorithmic}
\end{algorithm}
Training the substitute model with Eqn.~\ref{obj_function} is non-trivial because of the following two challenges. From the view of computational overhead, training the substitute model with all possible bitwidths is expensive and unnecessary in reality. We need to form a trade-off between the substitute model training overhead and the attack performance. From the view of batch normalization (BN) layers, BN is incapable of capturing the complex and largely divergent distributions of weights with different bitwidths, making it difficult for the model training process to converge. This problem has also been observed in other recent works~\cite{yoon2022bitwidth,wang2022removing}. We plot the BN parameters of 32-bit and 2-bit ResNet-34 in Fig.~\ref{fig:bn_visulization} to quantitatively illustrate this problem. In Fig.~\ref{fig:bn_visulization}, we show the channel-wise BN statistics of the 20th layers in the ResNet-34 ImageNet models obtained by full-precision training and quantized training. Each dot represents the running mean and variance of a channel. The full-precision DNN and low-bitwidth QNNs have very different weight distributions. Thus, we need to circumvent the BN incompatibility issue when training substitute models with Eqn.~\ref{obj_function}. The question then becomes the following: \textit{How can we circumvent the computational overhead and the BN incompatibility problem when training substitute models with multiple bitwidth objectives?}

To answer the above question, we propose training the substitute model with only two bitwidths (32-bit and 2-bit) to address the trade-off between the computational complexity and attack performance; that is,
\begin{equation}
\label{obj_function_2}
    \min_{w, \beta} \sum_{i}[\mathcal{L}(\overline{f}_{32}(x_i, w), y_i) + \mathcal{L}(\overline{f}_{2}(x_i, w, \beta), y_i)].
\end{equation}
We choose 32-bit and 2-bit settings because they represent two extreme cases, i.e., full-precision and low-precision cases. We do not use 1-bit QNN \cite{hubara2016binarized} because it is difficult to converge on ImageNet, and needs special tricks that are not compatible with other bitwidths. Regarding the BN incompatibility problem, although weight quantization is incompatible with the BN layer, applying the multiple bitwidths training objective only for activation quantization does not cause problems because activation quantization is performed after the BN layer (please see Fig.~\ref{fig:qaa} for reference). Therefore, we propose applying multiple bitwidths training to the activation quantization process only and keeping the weight quantized as 2-bit.

Alg.~\ref{alg:qaa} summarizes our QAA implementation. The adversary can resume from standardly pre-trained QNN to save the computational budget. In practice, fine-tuning for only one epoch is sufficient for enhancing the attack transferability to a large extent. The QAA trains the substitute model with quantized activation (line 5) and full-precision activation (line 9) alternatively while keeping the weight quantized. When generating adversarial examples, quantized activation (line 18) and full-precision activation (line 20) are also alternatively adopted, making the adversarial examples transferable across target models with different quantization bitwidths. The QAA is flexible and can be implemented using various quantization methods, including QAT~\cite{li2019additive} and PTQ~\cite{wei2022qdrop}. 

The implementation of the QAA based on PTQ is summarized in Alg.~\ref{alg:qaa_ptq}. Compared to the QAA implementation based on QAT, the QAA implementation based on PTQ is different in two ways. First, because PTQ does not train and quantize the model from scratch, the QAA implementation based on PTQ does not need model training or fine-tuning steps (there is no substitute model training step in Alg.~\ref{alg:qaa_ptq}); Second, instead of keeping weights quantized all the time, the QAA implementation based on PTQ applies multiple bitwidth inferences to both the weight and activation values (see the forward propagation step in Alg.~\ref{alg:qaa_ptq}). The PTQ version of the QAA does not require fine-tuning but is less powerful than the QAT version of the QAA.

\begin{algorithm}[t]
\caption{Quantization Aware Attack based on PTQ} \label{alg:qaa_ptq}
\begin{algorithmic}
\STATE \textbf{Require:} a pre-trained QNN $\overline{f}$ quantized with PTQ
\STATE \textbf{a. Adversarial example generation:}
\STATE \hspace{0.05cm} 1: \textbf{Initialize:} $useActQuant = True$, $x_{adv}=x$
\STATE \hspace{0.05cm} 2: \textbf{Repeat} for N iterations:
\STATE \hspace{0.05cm} 3: \hspace{0.5cm} $useActQuant$ = \textbf{not} $useActQuant$
\STATE \hspace{0.05cm} 4: \hspace{0.5cm} \textbf{if} $useActQuant$ \textbf{do}
\STATE \hspace{0.05cm} 5: \hspace{0.5cm}  \hspace{0.5cm} $output = \overline{f}(x_{adv}, \hat{w}, \hat{a})$
\STATE \hspace{0.05cm} 6: \hspace{0.5cm} \textbf{else do}
\STATE \hspace{0.05cm} 7: \hspace{0.5cm} \hspace{0.5cm} $output = \overline{f}(x_{adv}, w, a)$
\STATE \hspace{0.05cm} 8: \hspace{0.5cm} \textbf{end if}
\STATE \hspace{0.05cm} 9:  \hspace{0.5cm}  $loss = \mathcal{L}(output, y)$
\STATE 10: \hspace{0.5cm} update $x_{adv}$
\end{algorithmic}
\end{algorithm}

\subsection{Quantitative Analysis on QAA}
\label{subsec:quantitiative_analysis_on_qaa}
\begin{figure}[t]
    \centering
    \begin{minipage}{1\textwidth}
        \subfloat[]{\includegraphics[width=.24\textwidth]{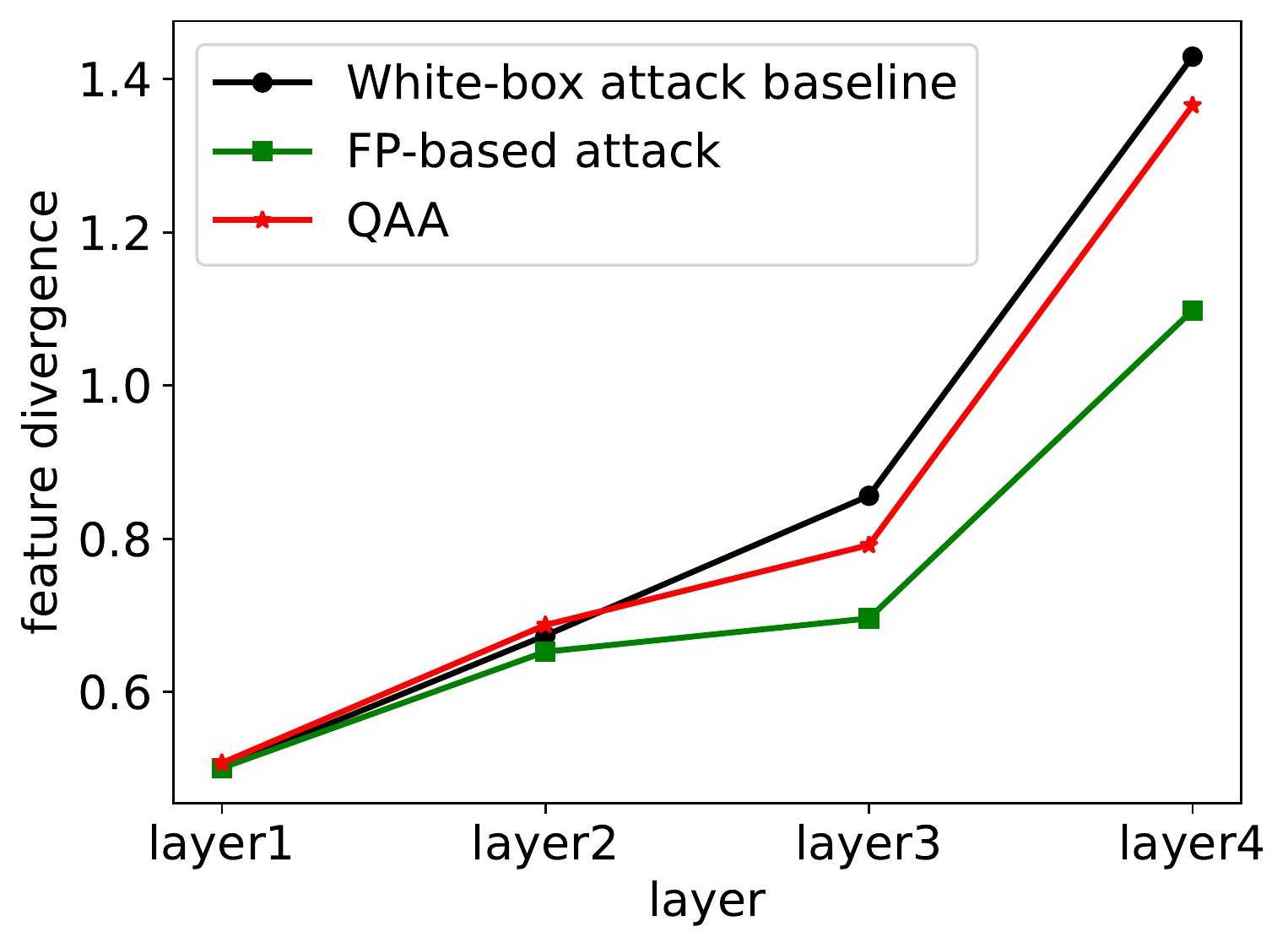}}
        \subfloat[]
        {\includegraphics[width=.24\textwidth]{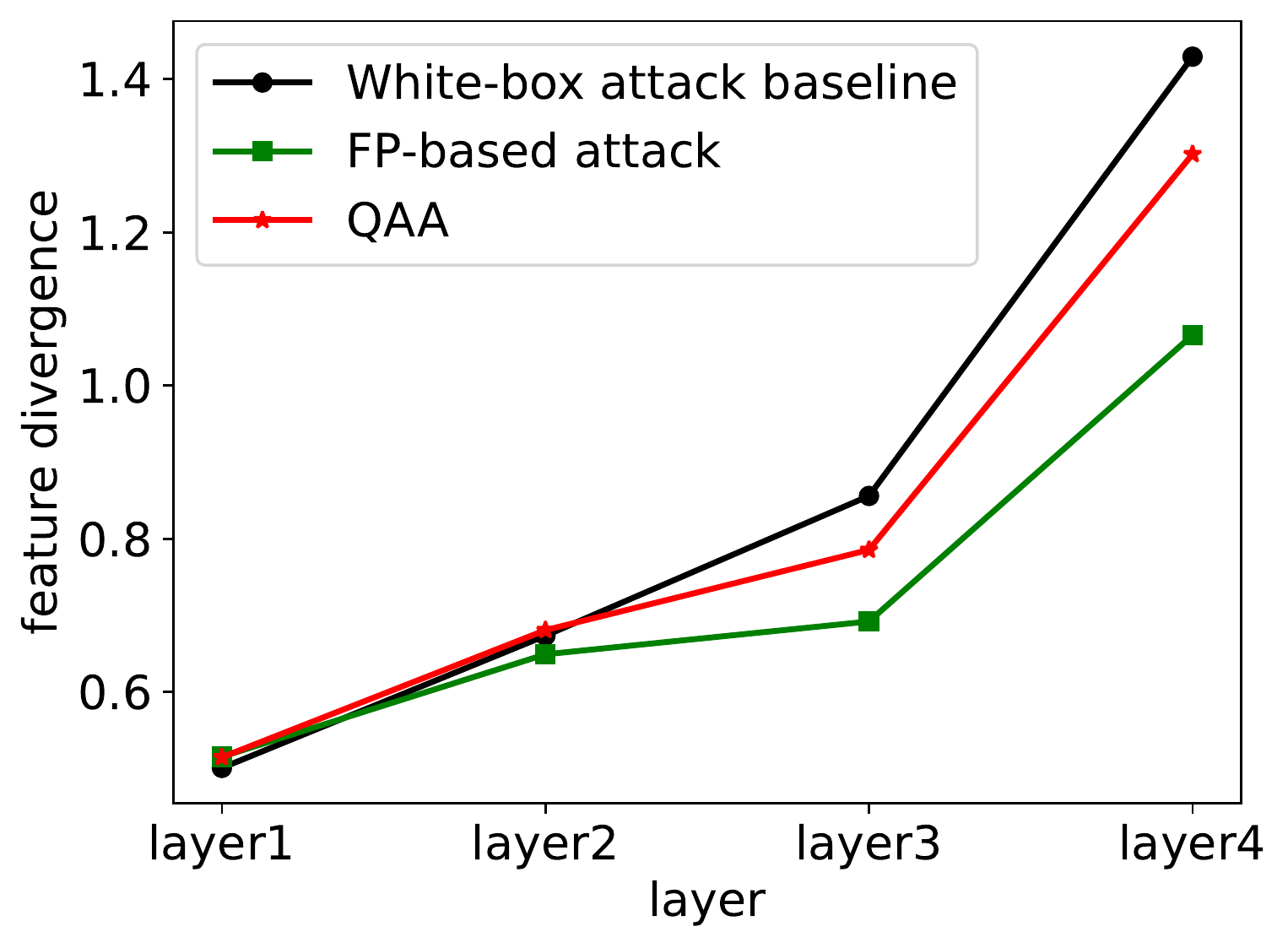}}
     \end{minipage}
     
     \begin{minipage}{1\textwidth}
        \subfloat[]
        {\includegraphics[width=.24\textwidth]           {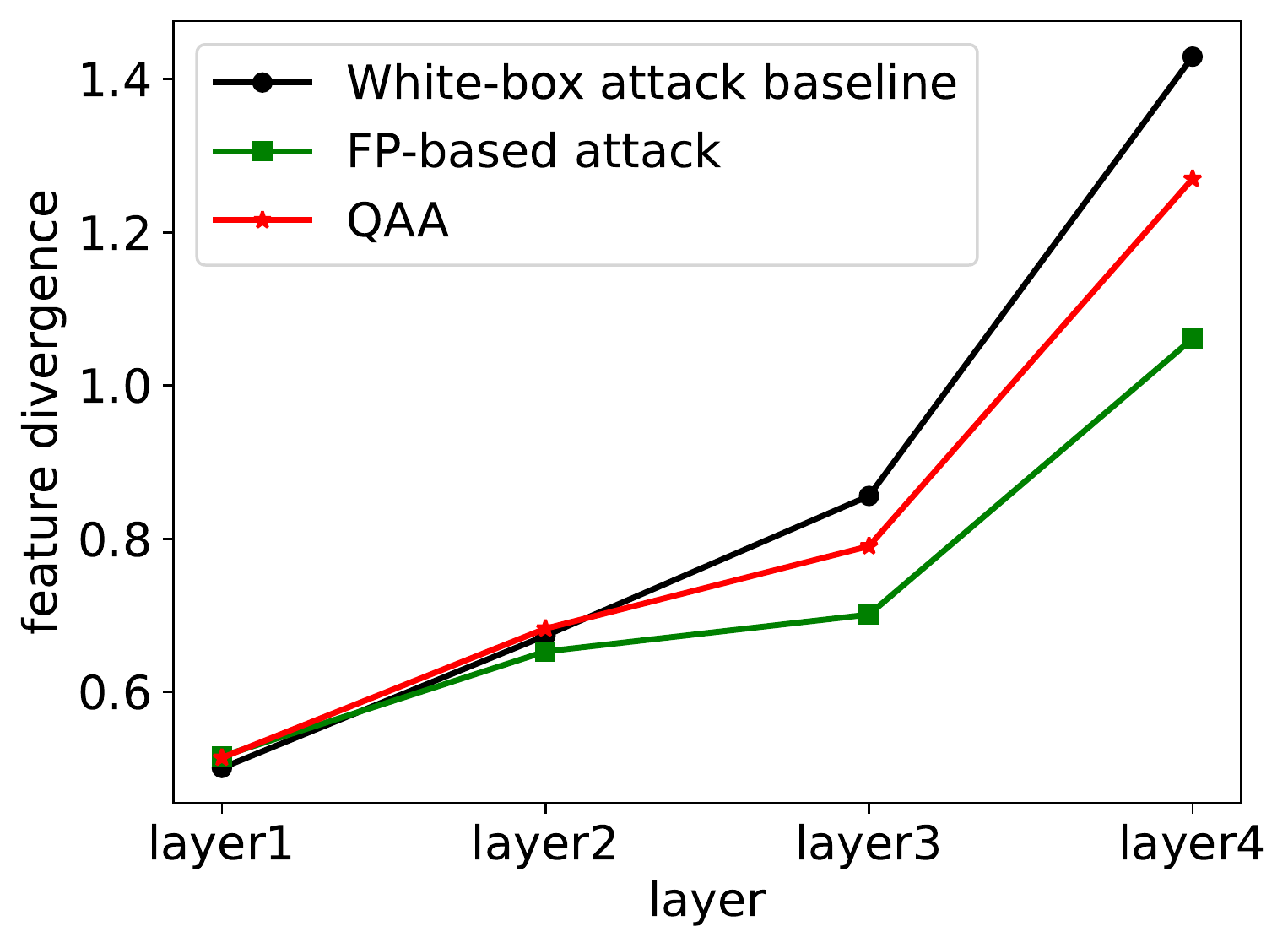}}
        \subfloat[]
        {\includegraphics[width=.24\textwidth]      
        {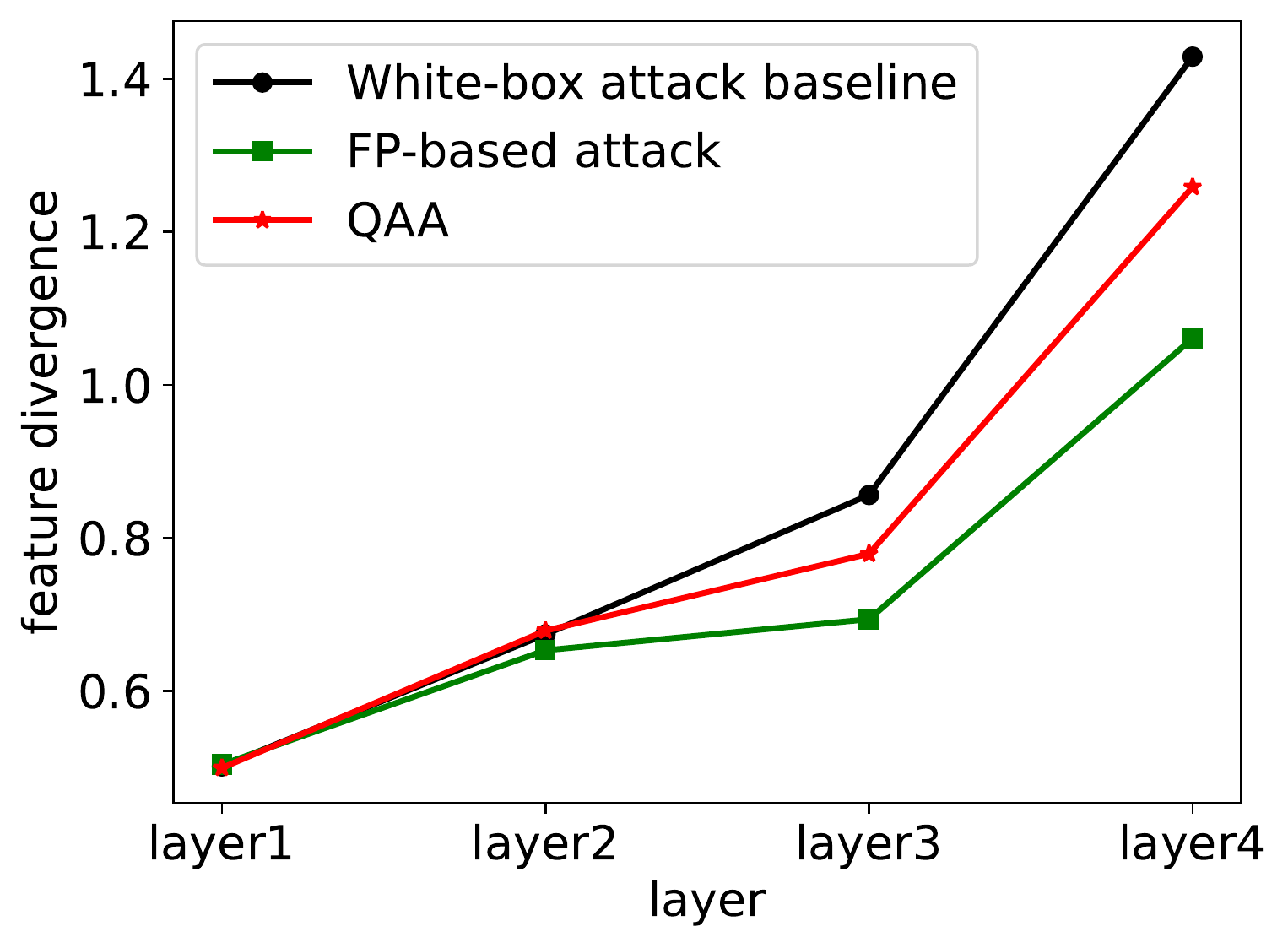}}
    \end{minipage}

    \caption{Illustration of how the QAA mitigates quantization shift. (a), (b), (c), and (d) show the feature divergence exhibited by 2, 3, 4, and 5-bit target QNNs, respectively.}
    \label{fig:quantization_shift}
\end{figure}

This subsection quantitatively analyzes whether the proposed QAA can mitigate the quantization shift and gradient misalignment issues mentioned above. To visualize the quantization shift phenomenon, we calculate the $k$-th layer feature divergence of the target model on an adversarial example and its corresponding clean example; that is,
\begin{equation}
    divergence_{k} = \frac{||f_k(x_{adv}) - f_k(x)||_2}{||f_k(x)||_2},
\end{equation}
where $f_k$ denotes the $k$-th layer of the target model $f$, $x$ is the clean example, $x_{adv}$ is the corresponding adversarial example, and $||\cdot||$ is the $l_2$-norm. We compare the feature divergence of the target model on adversarial examples generated with the QAA and the traditional attack based on the full-precision substitute model (FP-based attack). The model architecture is ResNet-34 for both the substitute models and target models, and the attack algorithm is MIM~\cite{dong2018boosting} with a perturbation budget of $\epsilon=16/255$. The results are presented in Fig.~\ref{fig:quantization_shift}. The feature divergence of the target QNN on the QAA attack is denoted by red lines, and the feature divergence on the FP-based attack is denoted by green lines. The black line denotes the white-box ideal case for reference. The utilized model is ResNet-34 for both the substitute and target models. The applied dataset is ImageNet. The attack is MIM with $l_{\infty}$-norm budget $\epsilon=16/255$. We can see that compared to the white-box baseline, the FP-based attack (denoted as the green line) substantially decreases the feature divergence, suggesting that the traditional FP-based attack suffers greatly from the quantization shift phenomenon. In comparison, the feature divergence of the QAA (denoted as the red line) decreases less. The above observation shows that the QAA is more robust to the quantization shift phenomenon and thus is more transferable across QNNs with different bitwidths.

\begin{figure}
    \centering
    \includegraphics[scale=0.6]{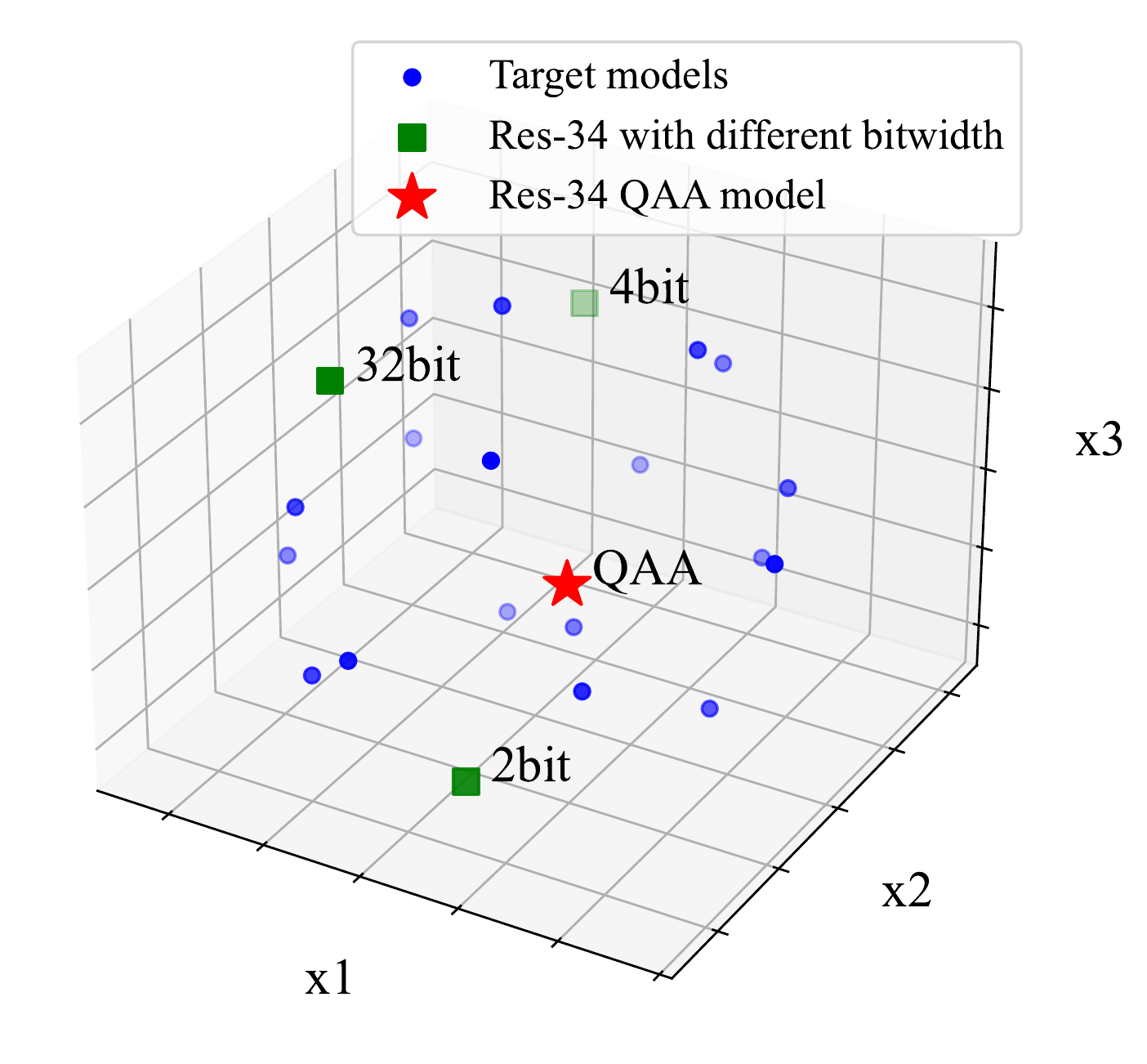}
    \captionsetup{justification=centering} 
    \caption{Visualization of the gradient alignment issue on ImageNet.}
    \label{fig:mds_visulization}
\end{figure}

To analyze the effectiveness of the QAA from the gradient misalignment perspective, we adopt the same method as that used in~\cite{bernhard2019impact,demontis2019adversarial}, calculating the gradient similarity between the target model and substitute model, that is, the cosine similarity of the gradient between the target model and substitute model,
\begin{equation}
    similarity = \frac{\nabla_x \mathcal{L}(f(x), y)^T \nabla_x\mathcal{L}(\overline{f}(x), y)}{||\nabla_x \mathcal{L}(f(x), y)||_2 \cdot
    ||\nabla_x \mathcal{L}(\overline{f}(x), y)||_2},
\end{equation}
where $T$ denotes the transpose of a matrix. Intuitively, a substitute model with higher gradient similarity to the target model enables the generation of more transferable adversarial examples. Specifically, we calculate the gradient similarity of each substitute model on 16 target models with different architectures and bitwidths (which are the same as those of the target models used in Sec.~\ref{sec:experiments}) and take the average. The 32-bit, 4-bit, 2-bit, and QAA-based ResNet-34 substitutes have average gradient similarities of 0.1180, 0.1186, 0.1196, and 0.1250, respectively. The maximum standard deviation of this experiment is $3.07\times 10^{-11}$, which shows that the QAA model has a statistically significant advantage. We draw Fig.~\ref{fig:mds_visulization} to visualize the gradient alignment. We calculate the gradient similarity between each model pair and define the distance between two models as 1 - gradient similarity. We use the multi-dimensional scaling (MDS) algorithm~\cite{kruskal1978multidimensional} to project the distance relationship into 3D space, as illustrated in Fig.~\ref{fig:mds_visulization}. Each point represents a model, and closer points indicate similar gradients. The QAA model is at the center of the target models; in other words, the QAA model has the shortest distances (in other words, highest gradient similarities) to these target models.

\section{Experiments}
\label{sec:experiments}
\subsection{Experimental Settings}

\textbf{Dataset.} We adopt ImageNet \cite{deng2009imagenet} and CIFAR-10 \cite{krizhevsky2009learning} to evaluate the QAA. On ImageNet, we follow a previous work~\cite{wang2021feature}, which conducted experiments on a subset of the ImageNet 2012 validation set containing 1,000 high-resolution images (299x299). On CIFAR-10, we use the test set to generate adversarial examples, which include 10,000 images (32x32). Please note that many previously developed transfer attacks are not effective on CIFAR-10 (see Tab.~\ref{tab:standard_benchmark_cifar10}), while the QAA is effective on both CIFAR-10 and ImageNet.

\textbf{Target Models.} We collect target models with various architectures, bitwidths, and defense methods to validate the generalizability of the QAA. \textcolor{black}{For ImageNet, we collect 21 target models, including seven standardly trained DNNs, five adversarially trained DNNs, and nine QNNs with bitwidths ranging from 2-bit to 8-bit. Here, we do not consider 1-bit target models on ImageNet because the adopted QDrop technique~\cite{wei2022qdrop} cannot achieve substantial accuracy under the 1-bit setting. For CIFAR-10, we collect 16 target models, including five standardly trained DNNs and 11 QNNs with bitwidths ranging from 1-bit to 8-bit.}

\textbf{Baseline Attacks.} On ImageNet, we combine the QAA with the SOTA transfer-based attacks designed for full-precision DNNs, including MIM~\cite{dong2018boosting}, CIM (=TI+DI+SI+MIM)~\cite{wang2021enhancing}, FIA~\cite{wang2021feature}, RPA~\cite{Zhang2022EnhancingTT}, Admix~\cite{wang2021admix}, and SSA~\cite{long2022frequency}. On CIFAR-10, we only combine the QAA with PGD, which is strong enough to largely outperform the SOTA transfer-based attacks. We adopt the $l_{\infty}$-norm to bound the adversarial perturbation and the $\epsilon$ values are $16/255$ and $8/255$ for ImageNet and CIFAR-10, respectively. In the tables in this section, we use \textbf{bold} numbers to denote higher transfer attack success rates than those of the baseline. The experimental results are in Tab.~\ref{tab:standard_benchmark_imagenet}, Tab.~\ref{tab:standard_benchmark_cifar10}, Tab.~\ref{tab:defense_benchmark_imagenet}, Tab.~\ref{tab:qnn_benchmark_imagenet}, and Tab.~\ref{tab:qnn_benchmark_cifar10} are repeated three times to calculate their maximum standard deviations. We list the maximum standard deviations in the caption, and they are much less than the improvement magnitude yielded by the QAA, illustrating that the QAA provides statistically significant improvements.

\textbf{Implementation Details.}
On ImageNet, we implement the QAA on Res-34 quantized with the QAT and Vgg-16 quantized with PTQ~\cite{wei2022qdrop}. On CIFAR-10, we implement the QAA based on Res-20 and Res-56 models quantized with QAT~\cite{li2019additive}. For both datasets, we fine-tune the 2-bit pre-trained QNNs for only one epoch. The batch size is 128, and the optimizer is SGD with a momentum of 0.9 and weight decay of 1e-4. The QAA implementation on PTQ does not need fine-tuning and can be directly used to generate adversarial examples after the QAA modification. The number of attack iterations is 10.

\begin{table*}[]
\centering
\caption{The Attack Success Rates (\%) Of QAA And Baseline Attacks Against Standardly Trained Target Models on ImageNet. 
The Maximum Standard Deviation is 1.1. The * denotes the white-box attack.}
\begin{tabular}{>{\cellcolor{white}}c|c|cccccccc}
\toprule
                         Substitute& Attack    & Inc-v3        & Inc-v4        & IncRes-v2     & Res-50        & Res-152       & Vgg-16         & Vgg-19        & Average       \\
\midrule
                         & MIM       & 43.8          & 42.4          & 34.2          & 76.2          & 63.1          & 76.3           & 71.6          & 58.2          \\
\rowcolor{gray!20}       & MIM+QAA   & \textbf{75.2} & \textbf{67.6} & \textbf{59.0} & \textbf{91.0} & \textbf{81.5} & \textbf{90.6}  & \textbf{88.7} & \textbf{79.1} \\
                         & CIM       & 88.0          & 82.2          & 77.5          & 96.0          & 92.6          & 93.0           & 93.3          & 88.9          \\
\rowcolor{gray!20}       & CIM+QAA   & \textbf{94.3} & \textbf{90.8} & \textbf{87.8} & \textbf{97.1} & \textbf{93.5} & \textbf{95.6}  & \textbf{95.6} & \textbf{93.5} \\
                         & FIA       & 77.5          & 76.9          & 63.5          & 95.2          & 89.1          & 91.2           & 91.6          & 83.6          \\
\rowcolor{gray!20}       & FIA+QAA   & \textbf{89.0} & \textbf{85.6} & \textbf{75.7} & \textbf{96.8} & \textbf{92.1} & \textbf{97.1}  & \textbf{95.6} & \textbf{90.3} \\
                         & RPA       & 76.8          & 77.3          & 63.8          & 95.0          & 89.1          & 92.5           & 91.8          & 83.8          \\
\rowcolor{gray!20}       & RPA+QAA   & \textbf{89.0} & \textbf{86.3} & \textbf{76.0} & \textbf{97.0} & \textbf{92.0} & \textbf{97.0}  & \textbf{96.1} & \textbf{90.5} \\
                         & Admix     & 54.7          & 51.5          & 40.8          & 85.4          & 73.9          & 83.9           & 80.9          & 67.3          \\
\rowcolor{gray!20}       & Admix+QAA & \textbf{74.6} & \textbf{67.4} & \textbf{59.0} & \textbf{92.0} & \textbf{81.4} & \textbf{92.3}  & \textbf{90.2} & \textbf{79.6} \\
                         & SSA       & 51.4          & 54.1          & 42.7          & 85.7          & 70.7          & 80.2           & 76.3          & 65.9          \\
\rowcolor{gray!20} \multirow{-12}{*}{Res-34} & SSA+QAA   & \textbf{66.6} & \textbf{63.0} & \textbf{49.3} & 85.4          & \textbf{73.8} & \textbf{84.5}  & \textbf{82.9} & \textbf{72.2} \\
\midrule
                         & MIM       & 33.2          & 39.3          & 27.8          & 54.9          & 37.8          & 99.9*           & 95.7          & 55.5          \\
\rowcolor{gray!20}       & MIM+QAA   & \textbf{39.5} & \textbf{46.5} & \textbf{31.6} & \textbf{65.2} & \textbf{44.0} & \textbf{100.0*} & \textbf{97.6} & \textbf{60.6} \\
                         & CIM       & 73.7          & 74.8          & 61.3          & 85.4          & 71.8          & 100.0*          & 99.6          & 80.9          \\
\rowcolor{gray!20}       & CIM+QAA   & \textbf{79.5} & \textbf{78.4} & \textbf{66.4} & \textbf{90.4} & \textbf{76.2} & \textbf{100.0*} & \textbf{99.4} & \textbf{84.3} \\
                         & FIA       & 56.4          & 66.5          & 46.0          & 82.3          & 62.5          & 100.0*          & 98.3          & 73.1          \\
\rowcolor{gray!20}       & FIA+QAA   & \textbf{65.3} & \textbf{72.6} & \textbf{52.5} & \textbf{86.7} & \textbf{70.3} & 99.9*  & \textbf{99.2} & \textbf{78.1} \\
                         & RPA       & 58.4          & 68.9          & 47.2          & 83.3          & 65.1          & 100.0*          & 98.6          & 74.5          \\
\rowcolor{gray!20}       & RPA+QAA   & \textbf{66.9} & \textbf{74.6} & \textbf{55.2} & \textbf{88.2} & \textbf{71.7} & \textbf{100.0*} & \textbf{99.6} & \textbf{79.5} \\
                         & Admix     & 29.0          & 37.4          & 21.4          & 59.0          & 40.4          & 99.1*           & 97.2          & 54.8          \\
\rowcolor{gray!20}       & Admix+QAA & \textbf{30.3} & \textbf{38.6} & \textbf{22.9} & \textbf{60.5} & \textbf{40.5} & \textbf{100.0*} & \textbf{98.7} & \textbf{55.9} \\
                         & SSA       & 38.4          & 43.7          & 31.8          & 60.7          & 44.3          & 100.0*          & 98.0          & 59.6          \\
\rowcolor{gray!20} \multirow{-12}{*}{Vgg-16} & SSA+QAA   & \textbf{39.3} & \textbf{44.3} & \textbf{32.3} & \textbf{61.4} & \textbf{44.6} & \textbf{100.0*} & \textbf{98.5} & \textbf{60.1} \\
\bottomrule
\end{tabular}
\label{tab:standard_benchmark_imagenet}
\end{table*}

\begin{table*}[]
\centering
\caption{The Attack Success Rates (\%) Of QAA And Baseline Attacks When Attacking Standardly Trained Target Models On CIFAR-10. The Maximum Standard Deviation Is 0.9.}
\begin{tabular}{>{\cellcolor{white}}c|c|cccccc}
\toprule
                        Substitute& Attack  & Res-18         & Res-50         & Vgg-19         & Dense-121      & MobileNet-v2   & Average        \\
\midrule
                        & PGD     & 39.44          & 43.66          & 71.73          & 41.31          & 90.06          & 57.24          \\
                        & MIM     & 49.54          & 54.14          & 73.19          & 51.04          & 86.69          & 62.92          \\
                        & CIM     & 40.20          & 42.41          & 45.50          & 41.70          & 56.61          & 45.28          \\
                        & FIA     & 40.30          & 43.44          & 68.49          & 41.89          & 86.04          & 56.03          \\
                        & RPA     & 39.94          & 42.09          & 67.42          & 40.70          & 85.40          & 55.11          \\
\rowcolor{gray!20} \multirow{-6}{*}{Res-20} & PGD+QAA & \textbf{61.99} & \textbf{64.77} & \textbf{87.62} & \textbf{63.09} & \textbf{96.15} & \textbf{74.72} \\
\midrule
                        & PGD     & 26.20          & 28.85          & 52.80          & 27.17          & 78.96          & 42.80          \\
                        & MIM     & 41.29          & 44.96          & 63.87          & 42.73          & 79.11          & 54.39          \\
                        & CIM     & 37.63          & 40.44          & 42.76          & 40.74          & 53.66          & 43.05          \\
                        & FIA     & 48.25          & 51.82          & 76.18          & 50.25          & 90.10          & 63.32          \\
                        & RPA     & 47.29          & 50.81          & 75.92          & 49.36          & 89.72          & 62.62          \\
\rowcolor{gray!20} \multirow{-6}{*}{Res-56} & PGD+QAA & \textbf{58.85} & \textbf{62.18} & \textbf{86.84} & \textbf{58.25} & \textbf{95.11} & \textbf{72.25} \\
\bottomrule
\end{tabular}
\label{tab:standard_benchmark_cifar10}
\end{table*}

\begin{table*}[]
\centering
\caption{The Attack Success Rates (\%) Against Adversarially Trained Models on ImageNet. The Maximum Standard Variation Is 0.7.}
\begin{tabular}{>{\cellcolor{white}}c|c|cccccc}
\toprule
                        Substitute & Attack    & Adv-Inc-v3    & Adv-IncRes-v2 & Ens3-Inc-v3   & Ens4-Inc-v3   & Ens-IncRes-v2 & Average       \\
\midrule
                         & MIM       & 27.7          & 23.0          & 26.5          & 24.3          & 15.2          & 23.3          \\
\rowcolor{gray!20}       & MIM+QAA   & \textbf{41.6} & \textbf{32.1} & \textbf{40.2} & \textbf{34.4} & \textbf{24.7} & \textbf{34.6} \\
                         & CIM       & 79.5          & 71.9          & 75.7          & 76.8          & 65.2          & 73.8          \\
\rowcolor{gray!20}       & CIM+QAA   & \textbf{88.9} & \textbf{81.4} & \textbf{85.0} & \textbf{83.6} & \textbf{74.1} & \textbf{82.6} \\
                         & FIA       & 60.2          & 49.2          & 59.0          & 50.5          & 38.8          & 51.5          \\
\rowcolor{gray!20}       & FIA+QAA   & \textbf{75.1} & \textbf{63.0} & \textbf{71.3} & \textbf{63.6} & \textbf{50.6} & \textbf{64.7} \\
                         & RPA       & 61.2          & 50.8          & 59.9          & 51.5          & 40.8          & 52.8          \\
\rowcolor{gray!20}       & RPA+QAA   & \textbf{76.6} & \textbf{65.9} & \textbf{72.1} & \textbf{64.6} & \textbf{52.2} & \textbf{66.3} \\
                         & Admix     & 34.6          & 29.7          & 33.8          & 28.8          & 20.3          & 29.4          \\
\rowcolor{gray!20}       & Admix+QAA & \textbf{54.3} & \textbf{43.5} & \textbf{50.9} & \textbf{42.7} & \textbf{33.3} & \textbf{44.9} \\
                         & SSA       & 36.5          & 24.7          & 32.0          & 29.5          & 17.7          & 28.1          \\
\rowcolor{gray!20} \multirow{-12}{*}{Res-34}& SSA+QAA   & \textbf{46.2} & \textbf{35.1} & \textbf{40.8} & \textbf{38.0} & \textbf{23.9} & \textbf{36.8} \\
\midrule
                         & MIM       & 19.1          & 13.5          & 18.3          & 15.7          & 9.8           & 15.3          \\
\rowcolor{gray!20}       & MIM+QAA   & \textbf{23.0} & \textbf{18.1} & \textbf{20.8} & \textbf{19.5} & \textbf{13.5} & \textbf{19.0} \\
                         & CIM       & 55.5          & 46.0          & 53.4          & 49.0          & 38.7          & 48.5          \\
\rowcolor{gray!20}       & CIM+QAA   & \textbf{62.4} & \textbf{50.9} & \textbf{59.6} & \textbf{55.7} & \textbf{44.6} & \textbf{54.6} \\
                         & FIA       & 35.5          & 24.5          & 31.3          & 26.8          & 20.0          & 27.6          \\
\rowcolor{gray!20}       & FIA+QAA   & \textbf{41.3} & \textbf{30.1} & \textbf{37.3} & \textbf{32.2} & \textbf{23.6} & \textbf{32.9} \\
                         & RPA       & 36.8          & 25.9          & 33.9          & 28.5          & 21.8          & 29.4          \\
\rowcolor{gray!20}       & RPA+QAA   & \textbf{43.9} & \textbf{33.6} & \textbf{39.2} & \textbf{32.1} & \textbf{25.1} & \textbf{34.8} \\
                         & Admix     & 16.1          & 11.1          & 15.4          & 13.8          & 8.8           & 13.0          \\
\rowcolor{gray!20}       & Admix+QAA & \textbf{22.5} & \textbf{15.0} & \textbf{19.6} & \textbf{18.8} & \textbf{11.6} & \textbf{17.5} \\
                         & SSA       & 21.7          & 14.8          & 20.0          & 18.4          & 11.1          & 17.2          \\
\rowcolor{gray!20} \multirow{-12}{*}{Vgg-16} & SSA+QAA   & \textbf{23.9} & \textbf{15.3} & \textbf{21.0} & \textbf{18.8} & \textbf{11.6} & \textbf{18.1}  \\
\bottomrule
\end{tabular}
\label{tab:defense_benchmark_imagenet}
\end{table*}

\begin{table*}[]
\centering
\caption{The Attack Success Rates (\%) Of QAA On QNN Target Models On CIFAR-10. The maximum standard deviation is 1.2.}
\begin{tabular}{c|c|cccccccccccc}
\toprule
                        \multirow{2}{*}{Substitute}& \multirow{2}{*}{Attack} & \multicolumn{2}{c}{NIN}         & \multicolumn{2}{c}{AlexNet}     & \multicolumn{2}{c}{Dense-121}   & 
                        \multicolumn{3}{c}{Res-20}                        & \multicolumn{2}{c}{MN-V2}       & \multirow{2}{*}{Average} \\
                        &                         & 8              & 1              & 8              & 1              & 8              & 4              & 8              & 4              & 2              & 8              & 4              &                          \\
\midrule
                        & PGD                     & 35.98          & 47.86          & 58.78          & 47.03          & 69.91          & 46.63          & 74.75          & 72.94          & 62.48          & 61.72          & 37.73          & 55.98                    \\
                        & MIM                     & 42.40          & 54.75          & 64.67          & 56.48          & 74.60          & 59.74          & 75.87          & 74.85          & 68.24          & 64.88          & 47.12          & 62.15                    \\
                        & FIA                     & 46.20          & 57.96          & 60.90          & 57.12          & 74.63          & 74.76          & 76.31          & 74.76          & 69.61          & 63.13          & 46.03          & 63.76                    \\
                        & CIM                     & 36.56          & 41.19          & 39.99          & 37.57          & 45.96          & 38.60          & 50.49          & 49.85          & 42.14          & 42.54          & 37.61          & 42.05                    \\
\rowcolor{gray!20} \cellcolor{white} \multirow{-5}{*}{Res-20}& QAA+PGD                 & \textbf{53.14} & \textbf{73.96} & \textbf{81.56} & \textbf{73.87} & \textbf{89.28} & \textbf{74.21} & \textbf{92.57} & \textbf{92.14} & \textbf{90.34} & \textbf{84.62} & \textbf{63.34} & \textbf{79.00}           \\
\midrule
                        & PGD                     & 30.24          & 36.29          & 39.84          & 34.02          & 53.64          & 32.96          & 59.81          & 56.16          & 49.55          & 46.88          & 27.81          & 42.47                    \\
                        & MIM                     & 38.43          & 48.39          & 53.99          & 48.54          & 65.10          & 50.86          & 68.42          & 64.49          & 60.77          & 57.20          & 41.28          & 54.32                    \\
                        & FIA                     & 48.74          & 62.09          & 66.83          & 61.29          & 79.03          & 61.40          & 79.37          & 78.40          & 75.88          & 69.90          & 52.03          & 66.81                    \\
                        & CIM                     & 33.64          & 37.04          & 36.64          & 34.52          & 41.06          & 34.99          & 46.49          & 45.55          & 40.10          & 39.67          & 34.56          & 38.57                    \\
\rowcolor{gray!20} \cellcolor{white} \multirow{-5}{*}{Res-56}& QAA+PGD                 & \textbf{53.96} & \textbf{77.37} & \textbf{82.49} & \textbf{79.79} & \textbf{91.46} & \textbf{77.24} & \textbf{93.53} & \textbf{94.70} & \textbf{91.31} & \textbf{86.20} & \textbf{66.96} & \textbf{81.36}     \\
\bottomrule
\end{tabular}
\label{tab:qnn_benchmark_cifar10}
\end{table*}

\begin{table*}[]
\centering
\caption{The Attack Success Rates (\%) Of Various Transfer Attacks Against QNNs On ImageNet. The maximum standard deviation is 0.5. The * Denotes the Attack within the same architecture.}
\begin{tabular}{c|c|cccccccccc}
\toprule
                         \multirow{2}{*}{Substitute}& \multirow{2}{*}{Attack} & \multicolumn{3}{c}{Res-50}                    & \multicolumn{3}{c}{Vgg-16}                       & \multicolumn{3}{c}{MobileNet-v2}              & \multirow{2}{*}{Average} \\
                         &                         & 8             & 4             & 2             & 8              & 4              & 2              & 8             & 4             & 3             &                          \\
\midrule
                         & MIM                     & 76.6          & 77.1          & 87.6          & 76.0           & 78.0           & 86.9           & 75.5          & 85.6          & 94.0          & 81.9                     \\
\rowcolor{gray!20}   \cellcolor{white}    & MIM+QAA                 & \textbf{91.1} & \textbf{92.9} & \textbf{97.0} & \textbf{89.7}  & \textbf{91.8}  & \textbf{95.0}  & \textbf{93.2} & \textbf{95.4} & \textbf{97.0} & \textbf{93.7}            \\
                         & CIM                     & 91.6          & 96.8          & 96.8          & 89.1           & 94.0           & 97.5           & 90.1          & 94.4          & 96.2          & 94.1                     \\
\rowcolor{gray!20}   \cellcolor{white}    & CIM+QAA                 & \textbf{97.1} & \textbf{97.6} & \textbf{98.6} & \textbf{95.7}  & \textbf{96.4}  & \textbf{98.6}  & \textbf{98.2} & \textbf{98.3} & \textbf{98.9} & \textbf{97.7}            \\
                         & FIA                     & 95.1          & 95.1          & 96.3          & 91.0           & 92.2           & 97.0           & 95.0          & 94.4          & 96.9          & 94.8                     \\
\rowcolor{gray!20}  \cellcolor{white}     & FIA+QAA                 & \textbf{96.7} & \textbf{97.1} & \textbf{99.0} & \textbf{96.9}  & \textbf{97.3}  & \textbf{98.9}  & \textbf{97.8} & \textbf{99.1} & \textbf{99.3} & \textbf{98.0}            \\
                         & RPA                     & 94.8          & 95.0          & 96.7          & 92.5           & 93.6           & 97.1           & 93.6          & 94.7          & 97.1          & 95.0                     \\
\rowcolor{gray!20}  \cellcolor{white}     & RPA+QAA                 & \textbf{96.5} & \textbf{97.0} & \textbf{98.8} & \textbf{96.2}  & \textbf{97.3}  & \textbf{98.2}  & \textbf{97.1} & \textbf{98.0} & \textbf{99.0} & \textbf{97.6}            \\
                         & Admix                   & 55.7          & 55.3          & 67.5          & 55.5           & 57.4           & 71.1           & 60.3          & 67.9          & 84.6          & 63.9                     \\
\rowcolor{gray!20} \cellcolor{white}       & Admix+QAA               & \textbf{92.0} & \textbf{93.0} & \textbf{98.2} & \textbf{92.2}  & \textbf{93.4}  & \textbf{96.5}  & \textbf{94.6} & \textbf{97.1} & \textbf{97.9} & \textbf{95.0}            \\
                         & SSA                     & 86.7          & 85.5          & 89.3          & 80.3           & 81.9           & 88.2           & 86.2          & 86.7          & 93.3          & 86.5                     \\
\rowcolor{gray!20} \cellcolor{white}  \multirow{-12}{*}{Res-34} & SSA+QAA                 & 85.9          & \textbf{87.3} & \textbf{93.5} & \textbf{84.4}  & \textbf{86.0}  & \textbf{93.5}  & \textbf{92.0} & \textbf{91.0} & \textbf{96.1} & \textbf{90.0}            \\
\midrule
                         & MIM                     & 54.5          & 53.9          & 76.3          & 99.9*           & 99.9*           & 99.8*           & 67.4          & 80.7          & 92.3          & 80.5                     \\
\rowcolor{gray!20}  \cellcolor{white}     & MIM+QAA                 & \textbf{65.6} & \textbf{68.5} & \textbf{84.2} & \textbf{100.0}* & \textbf{99.9}*  & \textbf{99.9}*  & \textbf{78.1} & \textbf{86.0} & \textbf{95.4} & \textbf{86.4}            \\
                         & CIM                     & 85.4          & 86.5          & 90.7          & 100.0*          & 100.0*          & 100.0*          & 90.7          & 92.1          & 95.8          & 93.5                     \\
\rowcolor{gray!20}   \cellcolor{white}    & CIM+QAA                 & \textbf{89.8} & \textbf{91.0} & \textbf{94.8} & \textbf{100.0}* & \textbf{100.0}* & \textbf{100.0}* & \textbf{94.8} & \textbf{94.9} & \textbf{97.0} & \textbf{95.8}            \\
                         & FIA                     & 82.1          & 82.5          & 91.4          & 100.0*          & 100.0*          & 99.6*           & 91.4          & 94..5         & 96.5          & 92.9                     \\
\rowcolor{gray!20}   \cellcolor{white}    & FIA+QAA                 & \textbf{86.6} & \textbf{87.3} & \textbf{93.8} & \textbf{99.9}*  & \textbf{99.9}*  & \textbf{99.9}*  & \textbf{94.2} & \textbf{96.2} & \textbf{97.4} & \textbf{95.0}            \\
                         & RPA                     & 83.9          & 83.3          & 91.8          & 100.0*          & 100.0*          & 99.6*           & 91.6          & 94.0          & 96.6          & 93.4                     \\
\rowcolor{gray!20}   \cellcolor{white}    & RPA+QAA                 & \textbf{88.4} & \textbf{89.2} & \textbf{95.0} & \textbf{100.0}* & \textbf{100.0}* & \textbf{99.9}*  & \textbf{95.0} & \textbf{96.7} & \textbf{97.7} & \textbf{95.8}            \\
                         & Admix                   & 60.7          & 59.8          & 76.4          & 99.1*           & 99.1*           & 99.1*           & 73.9          & 79.7          & 91.2          & 82.1                     \\
\rowcolor{gray!20} \cellcolor{white}      & Admix+QAA               & \textbf{61.0} & \textbf{61.9} & \textbf{79.3} & \textbf{100.0}* & \textbf{100.0}* & \textbf{100.0}* & \textbf{77.1} & \textbf{82.8} & \textbf{92.4} & \textbf{83.8}            \\
                         & SSA                     & 62.3          & 63.1          & 75.8          & 100.0*          & 100.0*          & 100.0*          & 76.0          & 79.1          & 88.5          & 82.8                     \\
\rowcolor{gray!20} \cellcolor{white} \multirow{-12}{*}{Vgg-16} & SSA+QAA                 & 61.6          & \textbf{64.6} & \textbf{77.6} & \textbf{100.0}* & \textbf{100.0}* & \textbf{100.0}* & \textbf{78.2} & \textbf{81.9} & \textbf{91.5} & \textbf{83.9} \\
\bottomrule
\end{tabular}
\label{tab:qnn_benchmark_imagenet}
\end{table*}

\subsection{Transferability of the QAA}
\textbf{Attacking Standardly Trained Models and Adversarially Trained Models.} Adversarial training is the SOTA adversarial defense method, which is selected as the target for evaluating the effectiveness of the QAA. The results are reported in Tab.~\ref{tab:standard_benchmark_imagenet} and Tab.~\ref{tab:defense_benchmark_imagenet}. We can see when the QAA is included, the baseline attack success rates against both standardly and adversarially trained target models are significantly enhanced. Specifically, QAA improves the attack success rates of MIM, CIM, FIA, RPA, Admix, and SSA on the Res-34 substitute model by 20.9\%, 4.6\%, 6.7\%, 6.7\%, 12,3\%, 6.3\% against standardly trained models; and 11.3\%, 8.8\%, 13.2\%, 13.5\%, 15.5\%, 8.7\% against adversarially trained models on average. Similar observations can be drawn from Tab.~\ref{tab:standard_benchmark_cifar10} concerning CIFAR-10. \textcolor{black}{In Tab.~\ref{tab:standard_benchmark_cifar10}, we can see that some previously developed transfer-based attacks (CIM, FIA, and RPA) do not outperform the traditional MIM and PGD attacks on CIFAR-10. This suggests that the effectiveness of these advanced attacks that were originally claimed on ImageNet may not generalize well to other (simpler) datasets such as CIFAR-10. This finding is consistent with that in ~\cite{zhu2022toward}. In comparison, the QAA largely outperforms the other transfer-based attacks on both CIFAR-10 and ImageNet, highlighting its good generalizability across different datasets. }

\textbf{Attacking QNNs with Different Bitwidths.} Tab.~\ref{tab:qnn_benchmark_imagenet} and Tab.~\ref{tab:qnn_benchmark_cifar10} report the attack success rates against QNNs with different architectures and bitwidths on ImageNet and CIFAR-10, respectively. The results show that the QAA achieves higher attack success rates than the baseline attacks, highlighting QAA's capability of improving the adversarial transferability across QNNs with unknown architectures and quantization bitwidths. We can see that on the Vgg-16 ImageNet substitute model, the QAA improves less when combined with Admix and SSA. This is because, on Vgg-16, we apply the QAA implementation based on PTQ, which does not fine-tune the substitute QNN but only ``self-ensemble'' it, which contradicts the intuition of Admix and SSA (Admix and SSA resort to image transformation to achieve ``self-ensembling''). However, for the Res-34 substitute model, even Admix and SSA improves substantially when combined with the QAA, highlighting the effectiveness of QAA fine-tuning.

\textbf{Remarks on the above transfer-based attack results.} Previous works~\cite{bernhard2019impact,sen2020empir,fu2021double} have investigated the attack transferability problem mainly from the defender's perspective and agree that adversarial examples transfer poorly across QNNs with the same architecture but different bitwidth, and this paper draws a different conclusion from the black-box attack perspective that extremely-low bitwidth QNN substitute models can be a powerful tool in generating adversarial examples across unknown model architectures and bitwidths. The experimental results in this section verify the above claim and our findings complement those of previous works.

\subsection{Ablation Study}
\label{subsec:ablation}
\begin{figure*}[]
    \centering
    \begin{minipage}{1\textwidth}
        \subfloat[]
        {\includegraphics[width=.32\textwidth]             {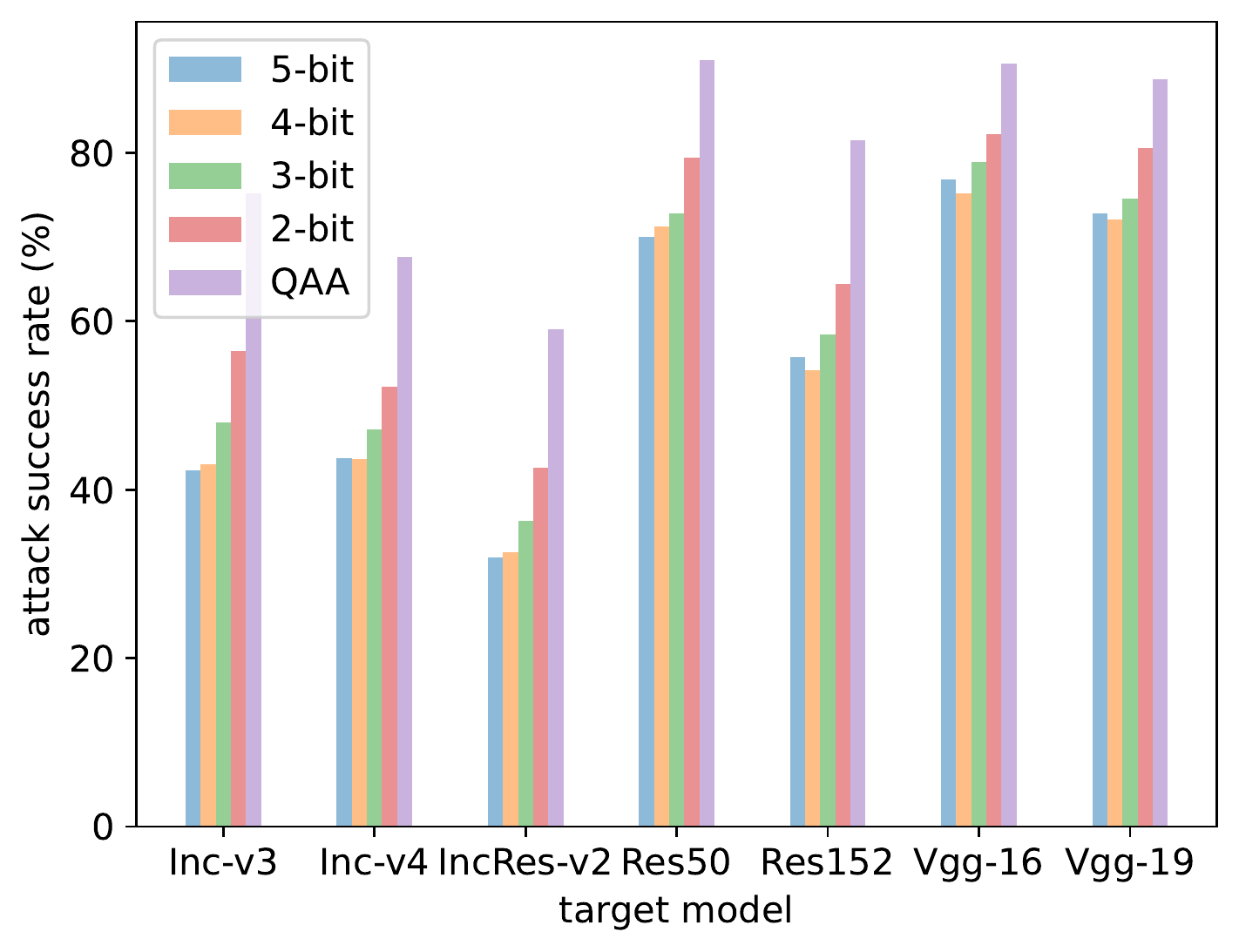}\label{fig:ablation_qnn}}
        \subfloat[]
        {\includegraphics[width=.32\textwidth]{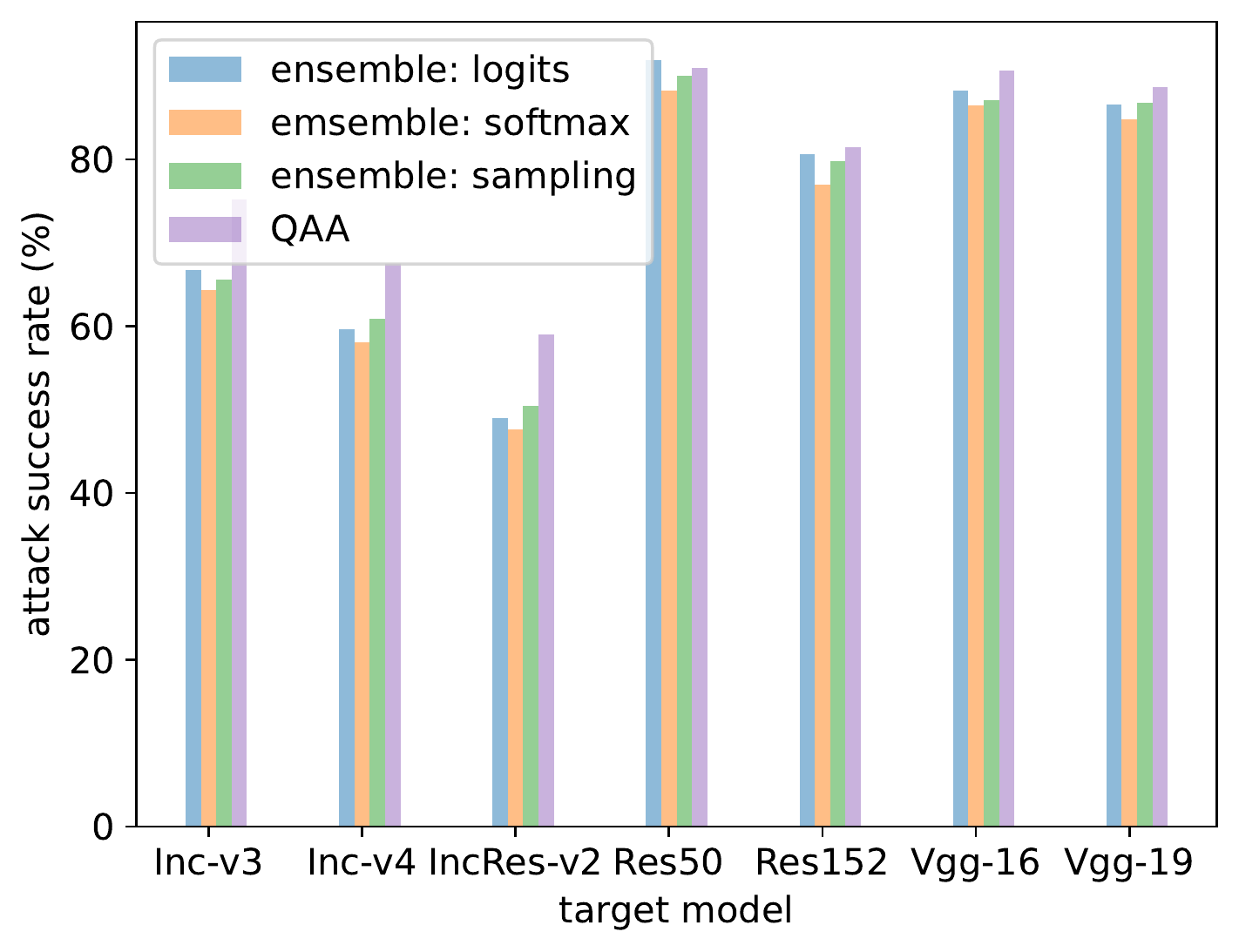}\label{fig:ablation_ensemble}}
        \subfloat[]
        {\includegraphics[width=.32\textwidth]{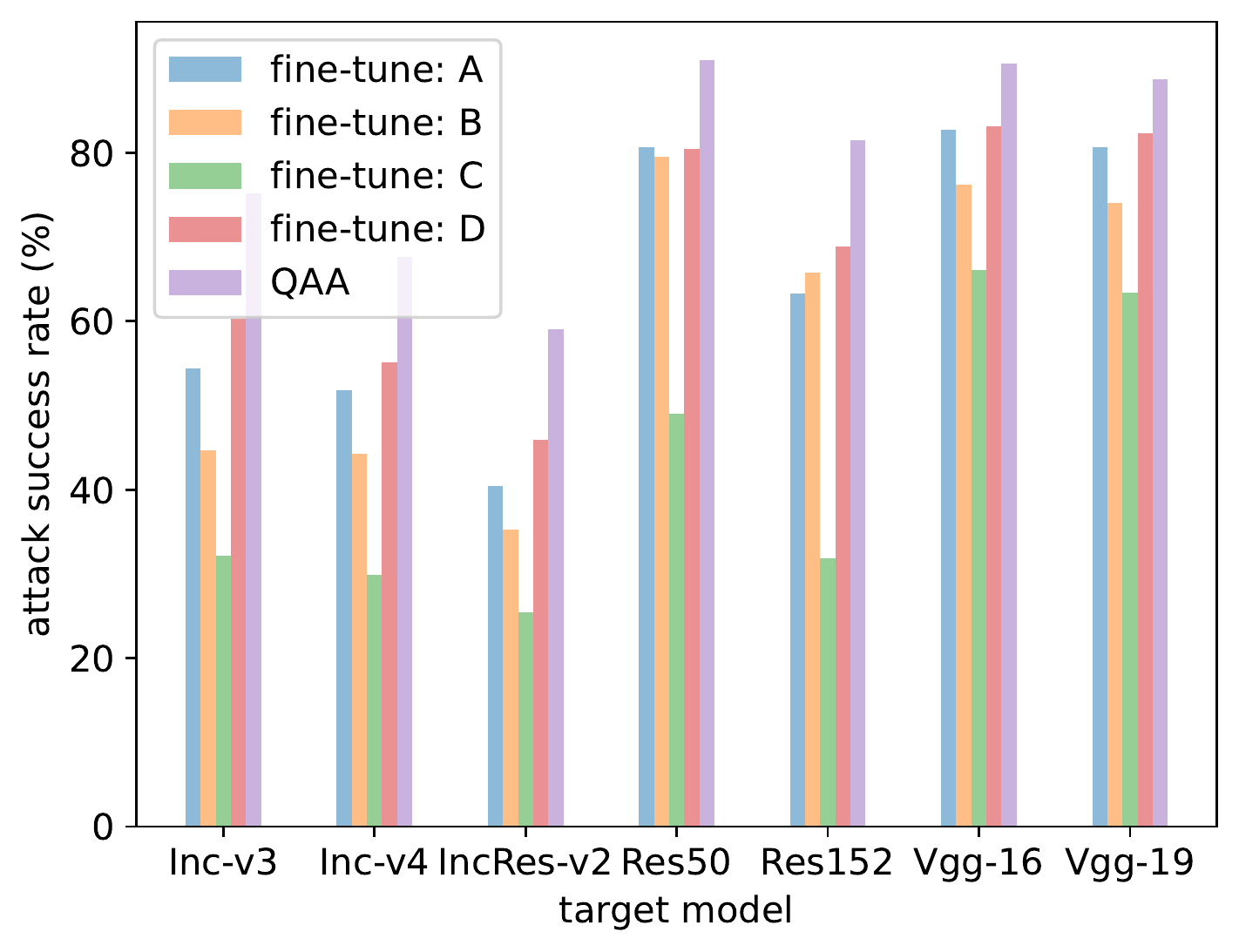}\label{fig:ablation_ft}}
    \end{minipage}
    \caption{Ablation study results. (a) shows the comparison results between the QAA and single-QNN based attacks (5, 4, 3, and 2-bit, respectively). (b) presents the comparison results between QAA and ensemble attacks with 32-bit and 2-bit models. We compare three different ensemble methods: logits, softmax, and sampling. (c) is the ablation studies on different fine-tuning objectives (A, B, C, D).
    }
\end{figure*}

\begin{table*}[]
\centering
\caption{Comparing the attack success rates and the training/storage overhead between LGV and QAA on standardly trained models.}
\begin{tabular}{c|c|c|c|c|c|c|c|c|c|c|c}
\toprule
                        Substitute& Attack  & Inc-v3          & Inc-v4          & IncRes-v2       & Res-50          & Res-152         & Vgg-16          & Vgg-19          & Average         & Training & Storage \\
\midrule
\multirow{3}{*}{Res-34} & QAA     & 94.3           & 90.8           & 87.8           & 97.1           & 93.5           & 95.6           & 95.6           & 93.5           & 1x       & 1x      \\
                        & LGV     & 93.1           & 90.4           & 86.0           & \textbf{98.4} & 94.8           & \textbf{97.9} & 97.1           & 94.0           & 10x      & 40x     \\
                        & LGV+QAA & \textbf{95.3} & \textbf{92.9} & \textbf{90.6} & 98.3           & \textbf{96.0} & 97.5           & \textbf{97.3} & \textbf{95.4} & 10x      & 40x  \\
\bottomrule
\end{tabular}
\label{tab:lgv_qaa_standard}
\end{table*}

\begin{table*}[]
\centering
\caption{Comparsion between LGV and QAA on adversarially trained models, and their training/storage computational overhead.}
\begin{tabular}{c|c|c|c|c|c|c|c|c|c}
\toprule
                        Substitute& Attack  & Adv-Inc-v3      & Adv-IncRes-v2   & Ens3-Inc-v3     & Ens4-Inc-v3     & Ens-IncRes-v2   & Average         & Training & Storage \\
\midrule
\multirow{3}{*}{Res-34} & QAA     & 88.9           & 81.4           & 85.0           & 83.6           & 74.1           & 82.6           & 1x       & 1x      \\
                        & LGV     & 86.8           & 80.4           & 86.0           & 84.6           & 74.0           & 82.4           & 10x      & 40x     \\
                        & LGV+QAA & \textbf{91.4} & \textbf{85.6} & \textbf{89.7} & \textbf{88.3} & \textbf{80.3} & \textbf{87.1} & 10x      & 40x \\
\bottomrule
\end{tabular}
\label{tab:lgv_qaa_defense}
\end{table*}

This section compares the QAA with other naive approaches to highlight its contribution. The dataset is ImageNet, the attack algorithm is MIM with $l_{\infty}$-norm $\epsilon=16/255$, and the substitute model architecture is ResNet-34 if not mentioned.

\textbf{Comparing QAA with QNN-based Attacks.} 
We compare the QAA with QNN-based attacks to verify that the QAA overcomes the shortcomings of substitute models with single bitwidths. For the QNN-based attacks, we use pre-trained QNNs (5, 4, 3, 2-bit) without any fine-tuning or other modifications as the substitute model. The results are presented in Fig.~\ref{fig:ablation_qnn}.
We can see that the QAA outperforms every compared baseline attack on every target model, verifying that the QAA overcomes the limitations of transfer-based attacks with single-bitwidth substitute models.

\textbf{Comparing the QAA with ensemble attacks}. To enhance the attack transferability against target QNNs, a naive and straightforward approach is to ensemble substitute QNNs with different bitwidths, which is denoted by ensemble attacks in this paper. Because QAA uses two quantization states (32-bit and 2-bit settings) in Eqn.~\ref{obj_function_2}, we ensemble 32-bit and 2-bit versions of ResNet-34 for a fair comparison. We have tried three commonly adopted ensemble techniques: ensembling at the logits layer (logits), ensembling at the softmax layer (softmax), and randomly sampling one model for each attack iteration (sampling). The results are presented in Fig.~\ref{fig:ablation_ensemble}. We can see that QAA outperforms every ensemble attack except on ResNet-50, on which the attack success rate of QAA is slightly lower than that of the logits attack.  Please note that the substitute model used in the experiment is also ResNet, which means that while QAA may not be better when attacking ResNet-50, it transfers better on totally different target architectures. Another advantage of the QAA is that its training, inference, and storage overheads are the same as those of a single QNN substitute model. In contrast, ensemble attacks multiply the computational overhead because they rely on training and running multiple substitute models. These experimental results demonstrate that the QAA provides further improvements and contributions compared to naive ensemble attacks. \textcolor{black}{We also compare the QAA with the ensemble attack LGV~\cite{gubri2022lgv} against standardly trained and adversarially trained target models in Tab.~\ref{tab:lgv_qaa_standard} and Tab.~\ref{tab:lgv_qaa_defense}, respectively. We can see that, compared to LGV, QAA achieves comparable transfer attack success rates with only 1/10 of the training overhead and 1/40 of the storage overhead. In addition, the QAA can also be combined with LGV to further enhance the attack transferability without increasing the computational overhead. The detailed description of LGV and the hyper-parameter settings of this experiment can be found in the \textit{Appendix}.}

\textbf{Comparing the QAA with other Training Objectives.} 
We compare the QAA with four other naive training objectives to highlight the contribution of the QAA. The hyper-parameters of the compared methods are the same as those of the QAA for a fair comparison. The compared methods include A. fine-tuning QNNs with only the original low bitwidth objective; B. fine-tuning 2-bit QNNs with 32-bit objective; C. fine-tuning 32-bit DNNs with Eqn.~\ref{obj_function_2}; and D. fine-tuning QNNs with 32-bit activation. The results in Fig.~\ref{fig:ablation_ft} show that each training method used in this experiment fails to achieve the same attack success rates as those of the QAA.

\subsection{Explaining the QAA}
This section explains why the QAA is effective from various perspectives. First, we take a closer look at the QAA to determine how the different states of QAA contribute to the final attack performance. Second, we explain the effectiveness of the QAA from the view of the loss landscape.
\begin{table*}[]
\centering
\caption{A closer look on QAA: testing the attack transferability of the QAA substitute model with fixed quantization states.}
\begin{tabular}{>{\cellcolor{white}}c|c|ccccccc}
\toprule
          Substitute& Bitwidth & Inc-v3          & Inc-v4          & IncRes-v2       & Res-50          & Res-152         & Vgg-16          & Vgg-19          \\
\midrule
                        & 32-bit          & 40.8           & 42.3           & 30.4           & 74.4           & 56.7           & 77.6           & 75.4           \\
                       & 32-bit QAA         & 48.7           & 46.1           & 37.9           & 72.1           & 54.3           & 79.1           & 77.0           \\
                        & 2-bit          & 47.8           & 46.6           & 34.6           & 73.6           & 53.1           & 81.4           & 80.7           \\
                  & 2-bit QAA          & 53.6           & 50.5           & 37.8           & 76.9           & 56.2           & 85.8           & 81.7           \\
\rowcolor{gray!20}  \multirow{-5}{*}{Res-34}   & QAA                & \textbf{66.0} & \textbf{59.1} & \textbf{45.4} & \textbf{85.3} & \textbf{67.5} & \textbf{91.2} & \textbf{87.6} \\
\midrule
                        & 32-bit           & 43.8           & 42.4           & 34.2           & 76.2           & 63.1           & 76.3           & 71.6           \\
                    & 32-bit QAA         & 59.0           & 54.7           & 46.2           & 80.7           & 64.9           & 80.1           & 79.0           \\
                        & 2-bit          & 56.5           & 52.2           & 42.6           & 79.4           & 64.4           & 82.2           & 80.6           \\
                      & 2-bit QAA          & 64.7           & 59.0           & 50.5           & 83.7           & 70.4           & 85.0           & 83.7           \\
\rowcolor{gray!20}   \multirow{-5}{*}{Res-18}                        & QAA                & \textbf{75.2} & \textbf{67.6} & \textbf{59.0} & \textbf{91.0} & \textbf{81.5} & \textbf{90.6} & \textbf{88.7} \\
\bottomrule
\end{tabular}
\label{tab:a_closer_look}
\end{table*}

\textbf{A closer look at the QAA.} As mentioned in Sec.~\ref{sec:method}, the QAA substitute model implements multiple quantization states when generating adversarial examples, including both the 32-bit activation state and the 2-bit activation state. It is natural to ask the following two questions. First, \textit{does each quantization state of the QAA substitute model improve compared to QNNs with the same bitwidth?} Second, \textit{does QAA improve upon each state it performs?} To answer these two questions, we conduct transfer-based attack experiments with standardly trained models on ImageNet, and the attack algorithm is MIM~\cite{dong2018boosting} with a $l_{\infty}$-norm budget of $\epsilon=16/255$. The results are shown in Tab.~\ref{tab:a_closer_look}. We can see that a single QAA quantization state improves upon single QNN substitute models (e.g., 2-bit QAA is better than 2-bit QNN), and the attack performance of the QAA is even greater than that of the QAA with a single state. The results are consistent across different target models and substitute models. Now we can conclude that the answer to the above two questions is yes. From this experiment, we can see that QAA not only improves the attack transferability of substitute models with single quantization states but also improves the transferability over these quantization states. Please note that these improvements come with a trivial computational cost, because the adversary only needs to fine-tune a pre-trained QNN for just one epoch to obtain a QAA substitute model, and the inference cost of the QAA model is the same as that of a single QNN model.

\begin{figure}
    \centering
    \includegraphics[scale=0.4]{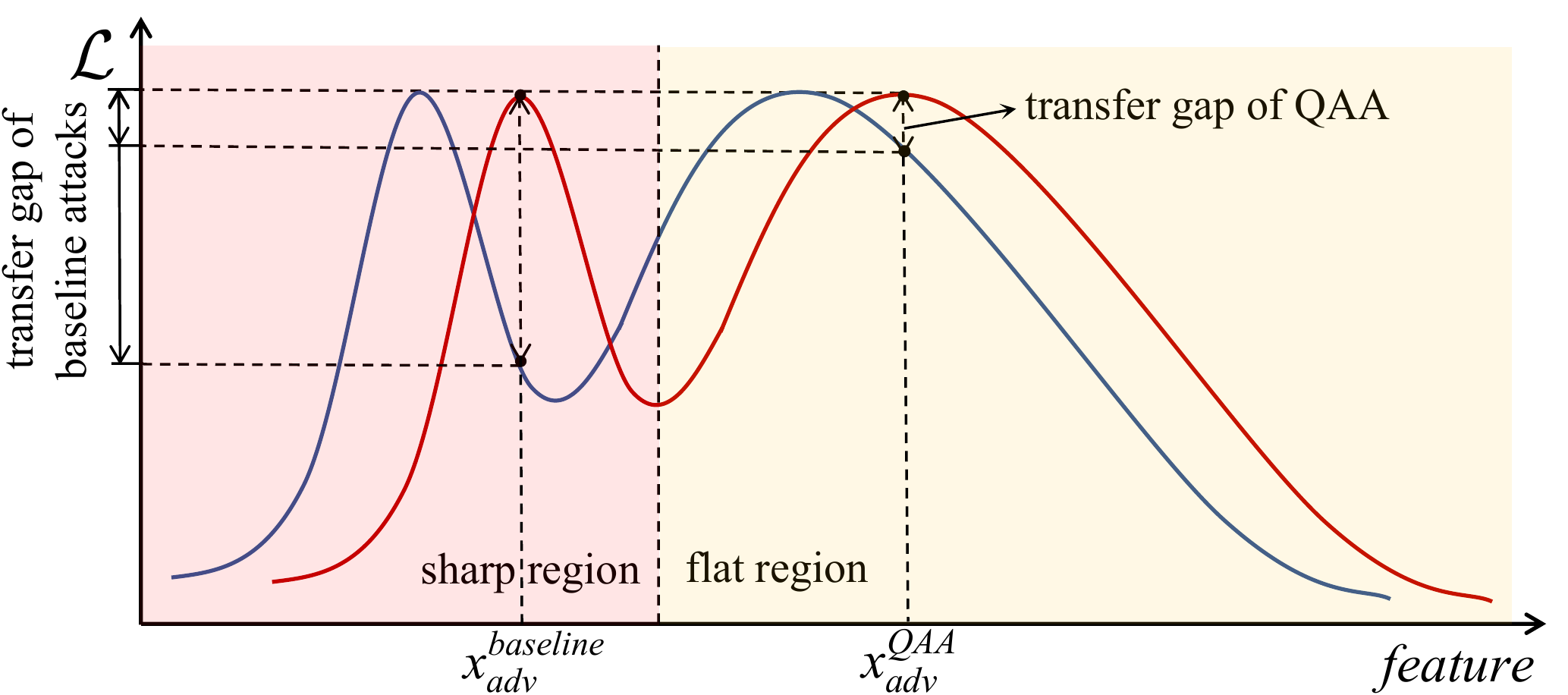}
    \caption{A conceptual sketch of flat and sharp regions in the feature space.}
    \label{fig:loss_landscape}
\end{figure}

\textbf{Explanation from the loss landscape view.} As mentioned in Section \ref{sec:method}, the QAA improves its transferability by overcoming quantization shift and gradient misalignment. This section provides a different explanation inspired by~\cite{gubri2022lgv}. The transferability between the two models can be explained by loss sharpness in both weight space and feature space. As stated in ~\cite{gubri2022lgv}, substitute models with flatter loss landscape in the weight space (in other words, lower sharpness in the weight space) enable the generation of adversarial examples in the flatter region of the loss landscape in the feature space. Adversarial examples in the flatter region of the feature space have better transferability.  Please see Fig.~\ref{fig:loss_landscape}, which is adapted from ~\cite{gubri2022lgv,keskar2016large} for reference and illustrates the loss landscape in the feature space. The Y-axis indicates the loss values, and the X-axis denotes the input features. The transfer gap of the adversarial examples in the flat region is significantly lower than that in the sharp region. The blue line marks the loss landscape of the target model, and the red line marks the loss landscape of the substitute model, which is slightly different from that of the target model. Adversarial examples falling on the flat region are more robust to slight landscape variations caused by the different architecture and quantization bitwidths between the substitute and target models; thus, they have better transferability. We believe that the QAA can promote the convergence of adversarial examples to flat regions. The following experimental results verify the above claim. 

Similar to~\cite{gubri2022lgv}, we calculate the loss sharpness in both the weight space and the feature space. The weight space sharpness of the substitute model is defined as
\begin{equation}
\label{eqn:sharpness_weight}
    \phi_{x, \overline{f}}^{w}(C_{\epsilon}) := \frac{\max_{\eta \in C_{\epsilon}}\overline{f}(x + \eta) - \overline{f}(x)}{1 + \overline{f}(x)} \times 100,
\end{equation}
where $C_{\epsilon}=\{z\in \mathbf{R}^p: -\epsilon \leq z_i \leq \epsilon, \forall i\in \{1,2,...,p\} \}$ is a constraint set with magnitude of $\epsilon$. As has been proven by previous work~\cite{keskar2016large}, Eqn.~\ref{eqn:sharpness_weight} is a good approximation of the magnitude of the eigenvalues $\nabla^{2}\overline{f}(x)$, and is used to characterize the sharpness of a model.
The feature space sharpness of the substitute model $\overline{f}$ around the adversarial example $x_{adv}$ is defined as
\begin{equation}
\label{eqn:sharpness_feature}
    \phi_{x, \overline{f}}^{x}(C_{\epsilon}) := \frac{\overline{f}(x_{adv}) - \min_{\eta \in C_{\epsilon}}\overline{f}(x_{adv} + \eta)}{1 + \min_{\eta \in C_{\epsilon}}\overline{f}(x_{adv} + \eta)} \times 100,
\end{equation}
where $C_{\epsilon}=\{z\in \mathbf{R}^p: -\epsilon \leq z_i \leq \epsilon, \forall i\in \{1,2,...,p\} \}$ is a constraint set with magnitude of $\epsilon$.
Intuitively, Eqn.~\ref{eqn:sharpness_feature} calculates the loss variation within a small ball around $x_{adv}$, and a lower sharpness suggests a flatter loss landscape in the feature space. We solve the above optimization problem with a projected gradient algorithm for 20 iterations and report the results with two values of $\epsilon$, $(5\cdot10^{-4}, 10^{-3})$ in Tab.~\ref{tab:sharpness}. Compared to transfer attacks using full-precision substitute models, the QAA consistently achieves lower sharpness values in both the weight space and the feature space, verifying that the QAA promotes adversarial examples to fall in the flat region and thus enhances the attack transferability. 

\begin{table}[]
\centering
\caption{The sharpness of each substitute model in both weight space and feature space with different $\epsilon$.
}
\begin{tabular}{c|ccccc}
\toprule
\multirow{2}{*}{Substitute}   & \multirow{2}{*}{Bitwidth} & \multicolumn{2}{c}{Sharpness (weight)} & \multicolumn{2}{c}{Sharpness (feature)} \\
                        &                           & $\epsilon$=5e-4               & $\epsilon$=1e-3              & $\epsilon$=5e-4               & $\epsilon$=1e-3               \\
\midrule
                        & 32                        & 60.8154                & 125.7775              & 71.8195                & 146.1035               \\
                        & 2  & 24.3069
                        & 52.1593 & 25.6317 & 60.5207 \\ 
\multirow{-3}{*}{Res-34} & QAA                       & 25.3686       & \textbf{51.9040}      & 27.4976       & \textbf{58.6584}       \\
\midrule
                        & 32                        & 76.9116                & 172.1912              & 91.3237                & 195.1994               \\
                        & 2 & 38.7024 & 86.0094 & 42.7450
                        & 101.1433 \\
\multirow{-3}{*}{Vgg-16} & QAA                       & \textbf{30.8196}       & \textbf{69.5630}      & \textbf{25.6498}       & \textbf{58.3942}      \\
\midrule
\end{tabular}
\label{tab:sharpness}
\end{table}

\section{Limitations and Future Work}

Despite the above achievements, this work still exhibits some limitations that deserve to be solved by future work.

\textbf{Mixed precision substitute model search.} Limited by the available computational resources, the mixed precision design of the QAA substitute model is straightforward. Although we perform ablation studies, it is unknown whether an even better mixed-precision substitute model can be obtained. Answering this question requires an exhaustive search process with many computational resources, which will be our future work. 

\textbf{Experiments on more models.} We test the effectiveness of the QAA on ResNets and Vggs with both the QAT and PTQ methods. Limited by the lack of available pre-trained QNN models (most of the previous model quantization works only implemented their methods on ResNets), we cannot provide experimental results for additional substitute models.

\textbf{More explanations of the attack transferability.} This paper explains the effectiveness of the QAA from three perspectives: quantization shift, gradient misalignment, and loss sharpness. Although these metrics are convincing with the support of previous works, we currently cannot provide more rigorous and insightful explanations because of the lack of explanatory works on DNNs and adversarial transferability. We leave this gap for future work.

\section{Potential Social Impact}
\textbf{Positive:} Strong transferability can benefit black-box applications of adversarial images for social goods, such as protecting user privacy~\cite{cherepanova2021lowkey,rajabi2021practicality,zhao2023adversarial,liu2019s,joon2017adversarial,larson2018pixel,liu2021pivoting}. In addition, the proposed approach can also motivate the community to design stronger defenses given our findings that even simple attacks can generate highly transferable adversarial images.

\textbf{Negative:} It is possible that our method may be susceptible to misuse by malicious entities aiming to compromise legitimate systems. However, we firmly believe that the substantial value our paper offers to the research community far surpasses any potential utility it might extend to malevolent actors.

\section{Conclusion}
In this paper, we propose a quantization aware attack (QAA) to mitigate the poor transferability of adversarial examples across target QNNs with different quantization bitwidths caused by quantization shift and gradient misalignment problems.
Extensive experimental results demonstrate the effectiveness of the QAA. The QAA lowers the barriers to attacking target QNNs with unknown quantization bitwidths and exposes the security risks of deploying QNNs in realistic scenarios. In the end, we explain the effectiveness of the QAA from the view of the loss landscape. Our findings remind the adversarial learning research community to pay attention to the vulnerability of QNNs but also raise new challenges for secure QNN design. 

\section{Acknowledgement}
This research is supported by the National Key Research and Development Program of China (2020AAA0107702), the National Natural Science Foundation of China (62376210, 62161160337, 62132011, U21B2018, U20A20177, U20B2049, 62206217, and 62006181), the Shaanxi Province Key Industry Innovation Program (2023-ZDLGY-38 and 2021ZDLGY01-02), the China Postdoctoral Science Foundation (2022M722530 and 2023T160512),
and the Fundamental Research Funds for the Central Universities under grant (xzy012022082, xtr052023004, and xtr022019002).

\appendix
\subsection{Settings for the Comparison with LGV}
LGV~\cite{gubri2022lgv} is a SOTA transfer-based attack that fine-tunes and ensembles the substitute model to enhance the attack transferability.  We implement LGV, QAA, and LGV+QAA on Res-34 with the ImageNet dataset, and the attack algorithm is CIM with perturbation budget $\epsilon=16/255$. 

The LGV re-implementation follows the original paper, which fine-tunes the pre-trained Res-34 for 10 epochs with a constant learning rate of 0.05 and collects 40 checkpoints for ensembling. In the attack generation stage, the collected checkpoints are randomly and non-repeatedly sampled. The training overhead of LGV is 10 times greater than that of the QAA, and the storage overhead is 40 times greater.

\begin{figure}[]
    \centering
    \includegraphics[scale=0.45]{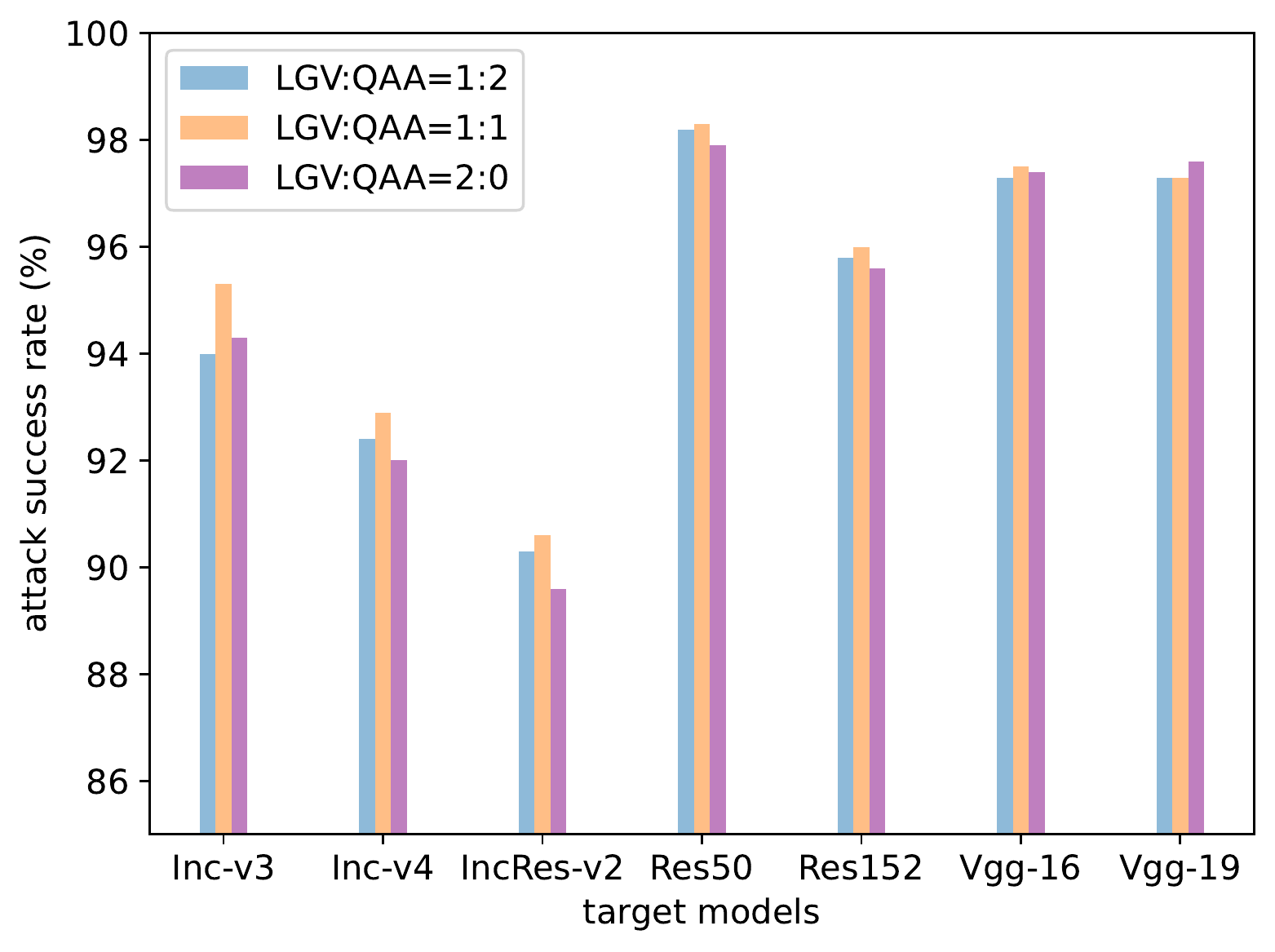}
    \caption{Hyper-parameter tuning for LGV+QAA.}
    \label{fig:lgv_qaa_tuning} 
\end{figure}

For LGV+QAA, we utilize either the QAA checkpoint or the LGV checkpoints in each attack iteration. To find the optimal sampling ratio, we perform multiple trials as shown in Fig.~\ref{fig:lgv_qaa_tuning} and set the sampling ratio to 1:1.

{
\bibliographystyle{IEEEtran}
\bibliography{reference}
}

\section{Biography Section}
\begin{IEEEbiography}[{\includegraphics[width=1in,height=1.25in,clip,keepaspectratio]{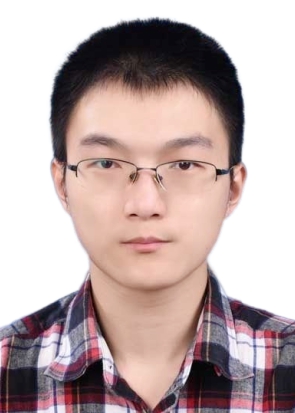}}]{Yulong Yang} received the B.Eng. degree in Computer Science and Engineering from Xi'an Jiaotong University in 2022, where he is currently pursuing the Ph.D. degree in Cyberspace Security with the School of Cyber Science and Engineering. His current research interests include adversarial machine learning and model compression.
\end{IEEEbiography}

\begin{IEEEbiography}
[{\includegraphics[width=1in,height=1.25in,clip,keepaspectratio]{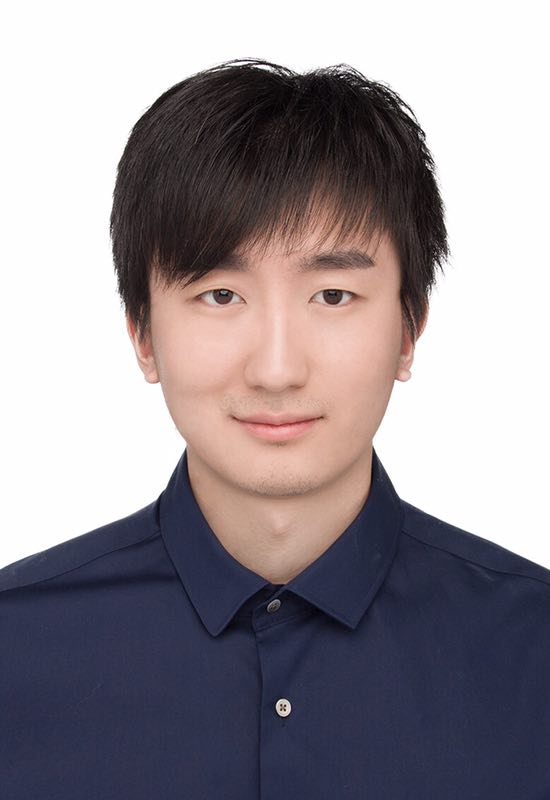}}]{Chenhao Lin} (Member, IEEE) received the B.Eng. degree in automation from Xi’an Jiaotong University in
2011, the M.Sc. degree in electrical engineering from Columbia University, in 2013, and the Ph.D. degree from The Hong Kong Polytechnic University, in 2018. He is currently a Research Fellow at the Xi’an Jiaotong University of China. His
research interests are in artificial intelligence security, identity authentication, biometrics, adversarial attack and robustness, and pattern recognition.
\end{IEEEbiography}

\begin{IEEEbiography}
[{\includegraphics[width=1in,height=1.25in,clip,keepaspectratio]{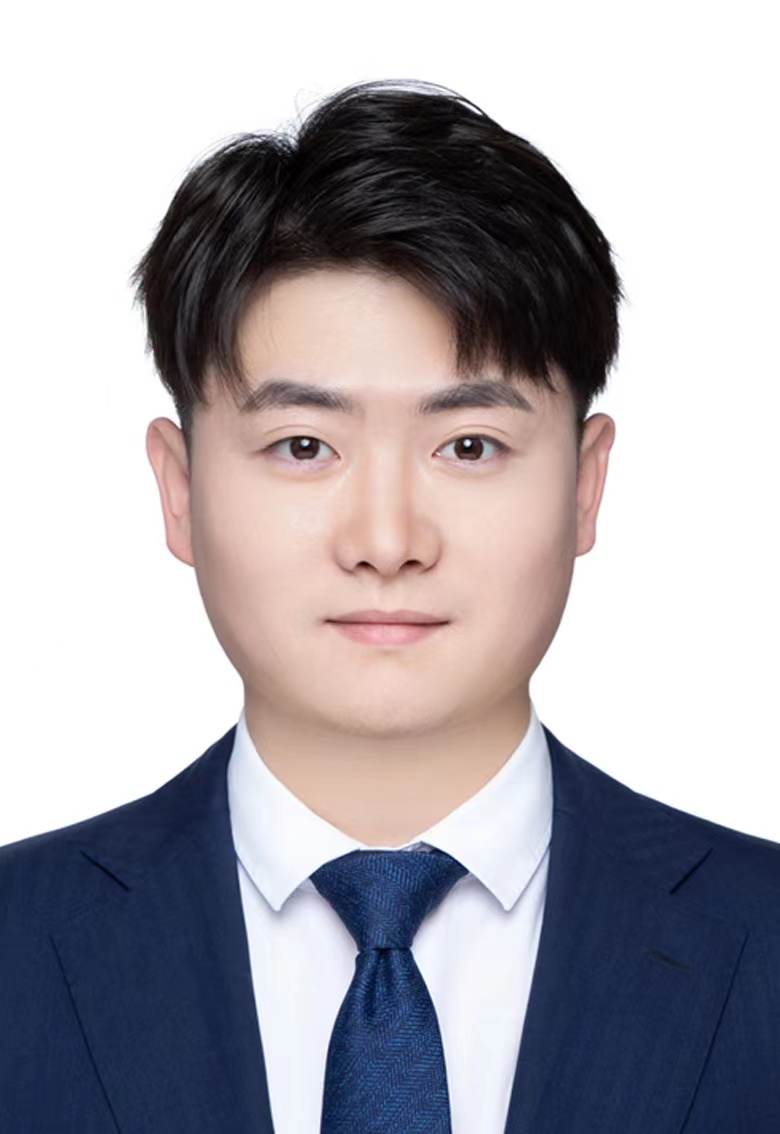}}] {Qian Li} (Member, IEEE) received the Ph.D. degree in computer science and technology from Xi’an Jiaotong University, China, in 2021. He is currently an Assistant Professor with the School of Cyber Science and Engineering, Xi’an Jiaotong University. His research interests include adversarial deep learning, artificial
intelligence security, and optimization of theory.
\end{IEEEbiography}

\begin{IEEEbiography}
[{\includegraphics[width=1in,height=1.25in,clip,keepaspectratio]{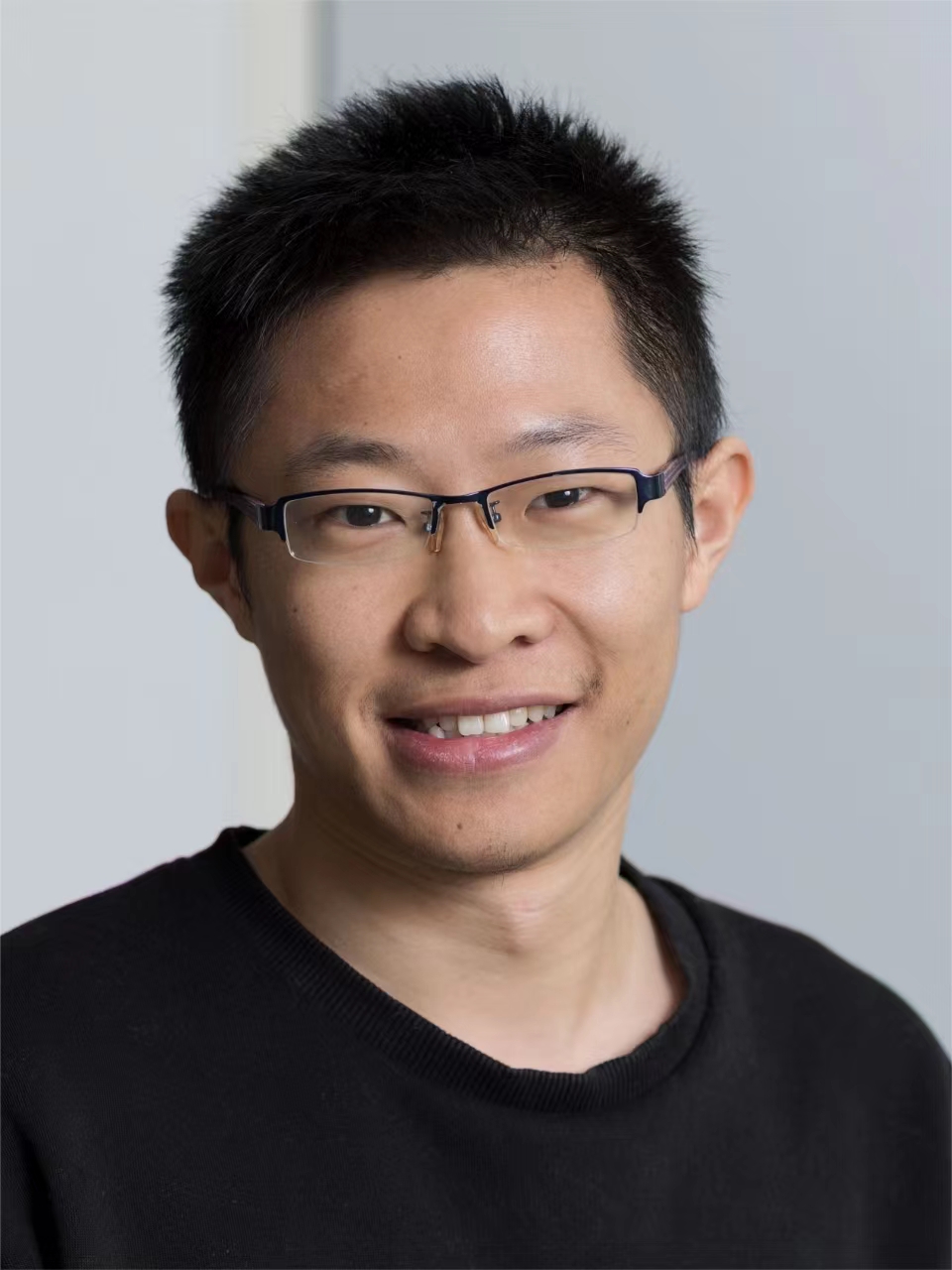}}]{Zhengyu Zhao} (Member, IEEE) received the Ph.D. degree from Radboud University, The Netherlands. He is currently an Associate Professor at Xi’an Jiaotong University, China. His general research interests include machine learning security and privacy. Most of his work has concentrated on security (e.g., adversarial examples and data poisoning) and privacy (e.g., membership inference) attacks against deep learning-based computer vision systems.
\end{IEEEbiography}

\begin{IEEEbiography}
[{\includegraphics[width=1in,height=1.25in,clip,keepaspectratio]{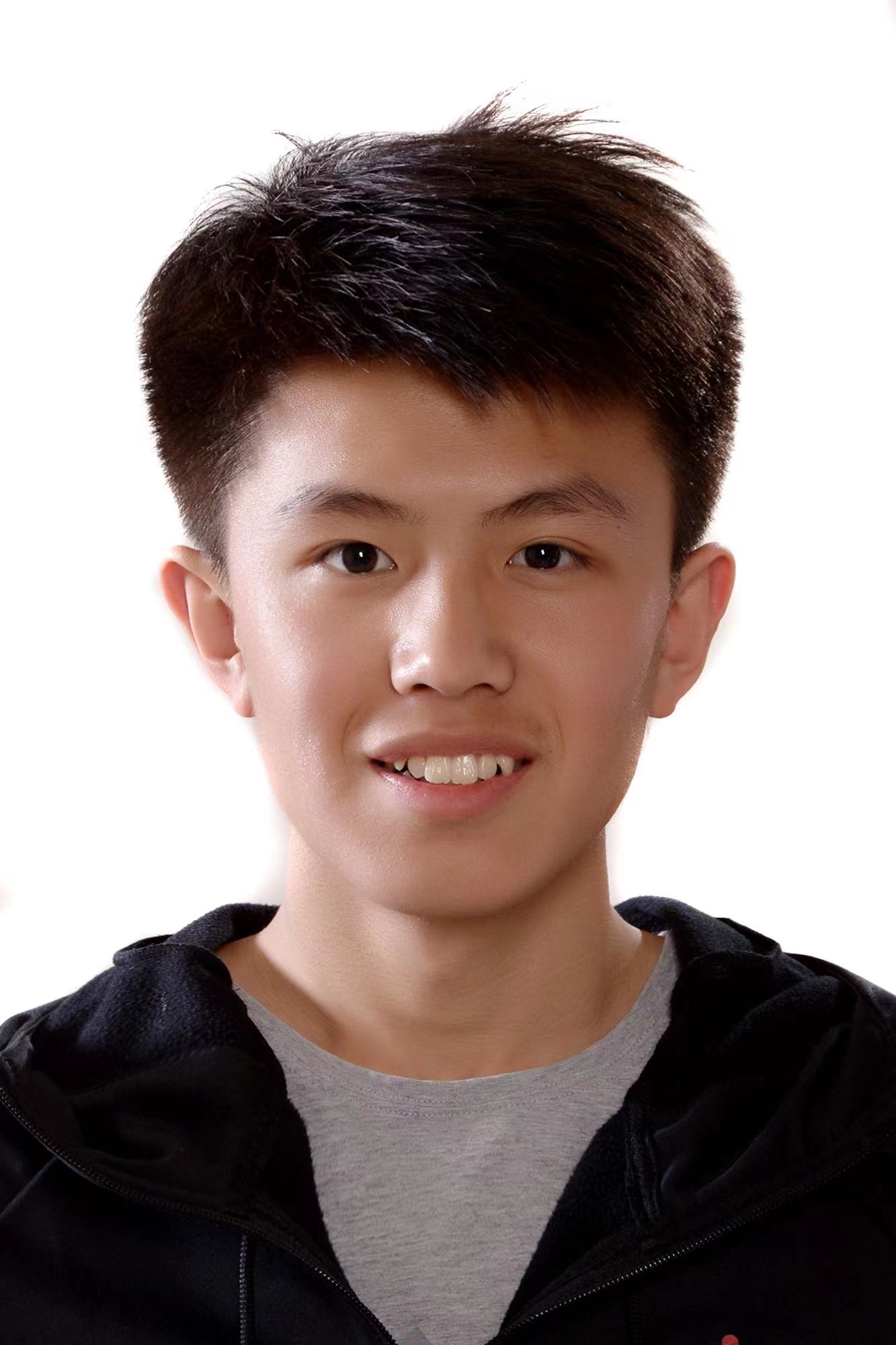}}]{Haoran Fan}
received the B.Eng. degree in Software Engineering from Dalian University of Technology in 2023 and is currently pursuing the M.Eng. degree in Software Engineering with the School of Software Engineering, Xi'an Jiaotong University. His current research interest is adversarial machine learning.
\end{IEEEbiography}

\begin{IEEEbiography}
[{\includegraphics[width=1in,height=1.25in,clip,keepaspectratio]{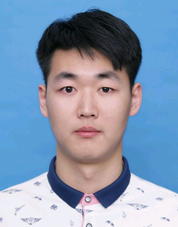}}]{Dawei Zhou} received the B.Eng. degree in telecommunications engineering from Xidian   University, Xian, China, in 2019, where he is currently pursuing the Ph.D. degree in information and communication engineering with the School of Telecommunications Engineering. His current research interests include computer vision and adversarial machine learning.
\end{IEEEbiography}

\begin{IEEEbiography}
[{\includegraphics[width=1in,height=1.25in,clip,keepaspectratio]{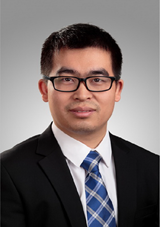}}]{Nannan Wang} (Member, IEEE) received the B.Sc. degree in information and computation science from the Xi’an University of Posts and Telecommunications in 2009 and the Ph.D. degree in information and telecommunications engineering from Xidian University in 2015.  He is currently a Professor with the State Key Laboratory of Integrated Services Networks, Xidian University. His current research interests include computer vision and machine learning.
\end{IEEEbiography}

\begin{IEEEbiography}
[{\includegraphics[width=1in,height=1.25in,clip,keepaspectratio]{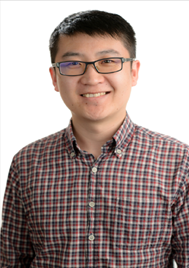}}]{Tongliang Liu} (Senior Member, IEEE) is currently a Lecturer and Director of the Trustworthy Machine Learning Lab with the School of Computer Science at the University of Sydney. He is broadly interested in the fields of trustworthy machine learning and its interdisciplinary applications, with a particular emphasis on learning with noisy labels, transfer learning, adversarial learning, unsupervised learning, and statistical deep learning theory.
\end{IEEEbiography}

\begin{IEEEbiography}
[{\includegraphics[width=1in,height=1.25in,clip,keepaspectratio]{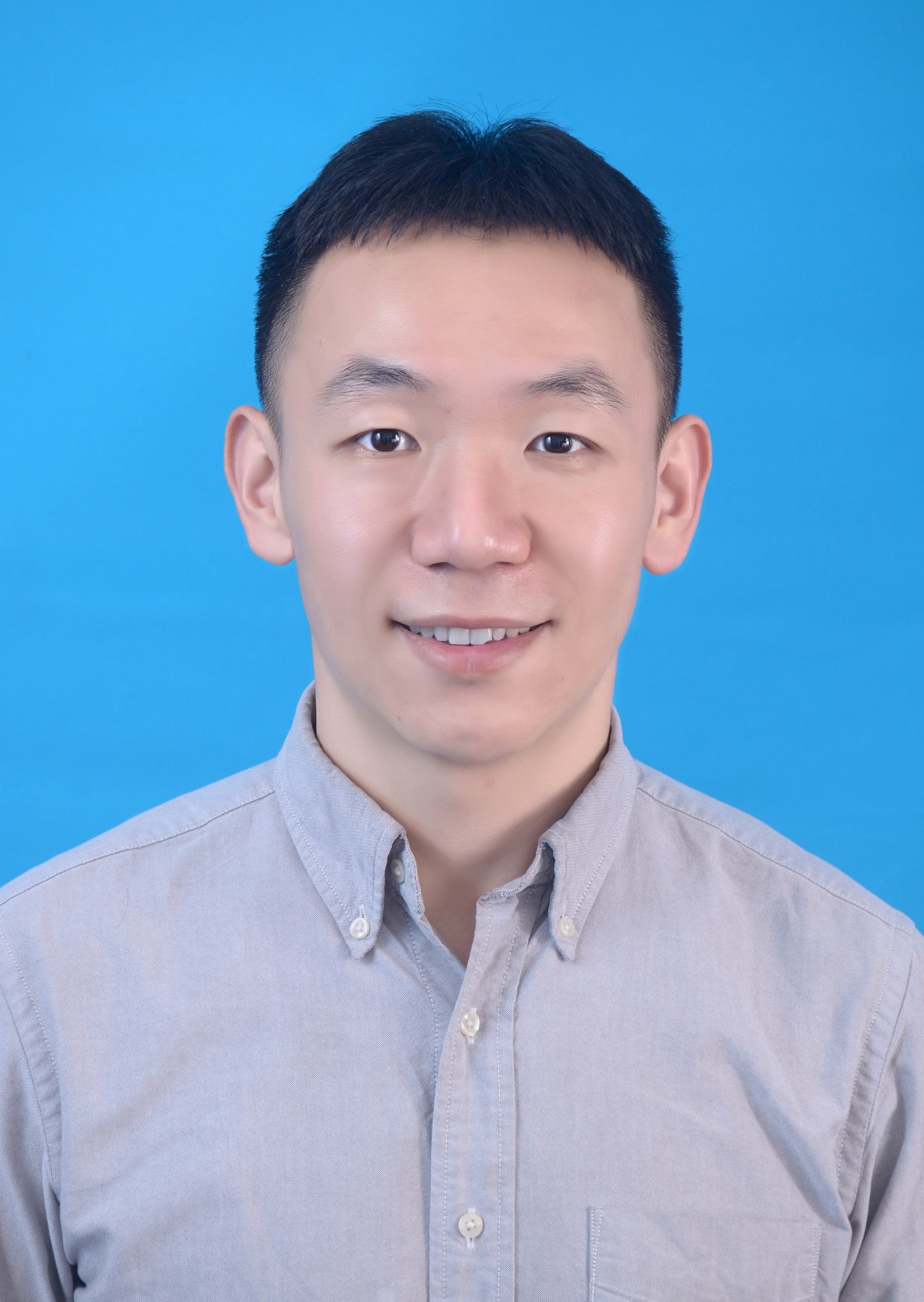}}]{Chao Shen} (Senior Member, IEEE)
received the B.S. degree in Automation from Xi’an Jiaotong University, China in 2007; and the Ph.D. degree in Control Theory and Control Engineering from Xi’an Jiaotong University, China in 2014. He is currently a
Professor in the Faculty of Electronic and Information Engineering, Xi’an Jiaotong University of China. His current research interests include AI Security, insider/intrusion detection, behavioral biometrics, and measurement and experimental
methodology.
\end{IEEEbiography}

\end{document}